\documentclass{aa}  
\usepackage{graphicx}
\usepackage{txfonts}
\usepackage{txfonts}
\usepackage{amsmath,amssymb}
\usepackage{natbib}
\usepackage[squaren,Gray]{SIunits}
\usepackage{color, colortbl}
\usepackage{array}
\usepackage{longtable,lscape}
\usepackage{multirow}

\newcommand {\micron}{\unit{}{\micro\meter}}

\begin{document} 

   \title{Optical polarized phase function of the HR\,4796A dust ring\thanks{The reduced images as FITS files presented in Fig. 1 are only available in electronic form at the CDS via anonymous ftp to cdsarc.u-strasbg.fr (130.79.128.5) or via http://cdsarc.u-strasbg.fr/ftp/vizier.submit//AA/2019/35363}}

   \author{
        J. Milli   \inst{1}
        \and N.Engler \inst{2}
         \and H.M. Schmid \inst{2}
         \and J. Olofsson \inst{3,4,5}
        \and F. M\'enard \inst{6}
         \and Q. Kral \inst{7}
         \and A. Boccaletti \inst{7}
         \and P. Th\'ebault \inst{7}
         \and E. Choquet \inst{8}
        \and D. Mouillet   \inst{6}
        \and    A.-M. Lagrange \inst{6} 
        \and J.-C. Augereau \inst{6}
         \and C. Pinte \inst{6}
         \and G. Chauvin \inst{8,9}
         \and C. Dominik \inst{10}
         \and C. Perrot \inst{4}
         \and A. Zurlo \inst{8,11,12}
         \and T. Henning \inst{3}
         \and J.-L. Beuzit \inst{8}
         \and H. Avenhaus \inst{3}
        \and A. Bazzon \inst{2}
        \and T. Moulin \inst{6} 
        \and M. Llored \inst{12}
        \and O. Moeller-Nilsson \inst{3}
        \and R. Roelfsema \inst{13}
        \and J. Pragt \inst{13}
          }

   \institute{
         European Southern Observatory (ESO), Alonso de C\'ordova 3107, Vitacura, Casilla 19001, Santiago, Chile\\
         \email{jmilli@eso.org} 
        \and
        Institute for Particle Physics and Astrophysics, ETH Zurich, Wolfgang-Pauli-Strasse 27, 8093 Zurich, Switzerland  
          \and 
          Max Planck Institute for Astronomy, K\"onigstuhl 17, D-69117 Heidelberg, Germany 
        \and
        Instituto de F\'isica y Astronom\'ia, Facultad de Ciencias, Universidad de Valpara\'iso, Av. Gran Breta\~na 1111, Playa Ancha, Valpara\'iso, Chile 
         \and
         N\'ucleo Milenio Formaci\'on Planetaria - NPF, Universidad de Valpara\'iso, Av. Gran Breta\~na 1111, Valpara\'iso, Chile 
         \and
        Univ. Grenoble Alpes, CNRS, IPAG, F-38000 Grenoble, France. 
         \and
          LESIA, Observatoire de Paris, Universit\'e PSL, CNRS, Sorbonne Universit\'e, Univ. Paris Diderot, Sorbonne Paris Cit\'e, 5 place Jules Janssen, 92195 Meudon, France 
          \and 
          Aix Marseille Universit\'e, CNRS, CNES,  LAM, Marseille, France 
          \and
           UMI-FCA, CNRS/INSU France (UMI 3386), and Departamento de Astronomia, Universidad de Chile, Casilla 36-D Santiago, Chile 
          \and
          Anton Pannekoek Institute for Astronomy, Science Park 904, NL-1098 XH Amsterdam, The Netherlands 
          \and 
          N\'ucleo de Astronom\'ia, Facultad de Ingenier\'ia y Ciencias, Universidad Diego Portales, Av. Ejercito 441, Santiago, Chile 
          \and
        Escuela de Ingenier\'ia Industrial, Facultad de Ingenier\'ia y Ciencias, Universidad Diego Portales, Av. Ejercito 441, Santiago, Chile 
          \and
          NOVA Optical Infrared Instrumentation Group, Oude Hoogeveensedijk 4, 7991 PD Dwingeloo, The Netherlands 
             }


   \date{Received 26 February 2019; accepted 28 April 2019}

 
  \abstract
   {
  The scattering properties of the dust originating from debris discs are still poorly known. The analysis of scattered light is however a powerful remote-sensing tool to understand the physical properties of dust particles orbiting other stars. Scattered light is indeed widely used to characterise the properties of cometary dust in the solar system.  
   }
   {
   We aim to measure the morphology and scattering properties of the dust from the debris ring around HR\,4796\,A in polarised optical light.
   }
   {
   We obtained high-contrast polarimetric images of HR\,4796\,A in the wavelength range 600-900nm with the SPHERE / ZIMPOL instrument on the Very Large Telescope.
   }
   {
   We  measured for the first time the polarised phase function of the dust in a debris system over a wide range of scattering angles in the optical. We confirm that it is incompatible with dust particles being compact spheres under the assumption of the Mie theory, and propose alternative scenarios compatible with the observations, such as particles with irregular surface roughness or aggregate particles. 
   }
   {}



   \maketitle
%

\section{Introduction}

Debris discs are a common outcome of stellar and planetary evolution, with a detection rate above 20\% for A-type stars \cite[e.g. ][]{Matthews2014}. Mostly detected through their infrared (IR) excess, they consist of one or several belts of  approximately kilometre-sized planetesimals, producing smaller debris in a collisional cascade, the smallest particles being blown out of the system by the radiation pressure of the central star. The dust is constantly replenished over several hundred million years through collisions \cite[see][for recent reviews]{Hughes2018,Kral2018_prospective}. Progress in high-angular resolution imaging techniques, both in the sub-millimetre and in the optical/near-infrared (NIR) regime, reveal the morphology of those belts in great detail, opening up new perspectives to characterise the properties and distribution of the emitting particles. 

This is particularly true for HR\,4796\,A, an A-type star  located at $71.91\pm0.70$ pc \citep{Gaia2018} with an estimated age of $10\pm3$ Myr \citep{Bell2015}. This star hosts one of the brightest debris discs, with a fractional luminosity reaching 0.5\% of the total system luminosity \citep{Moor2006}. For this reason, it has been observed from ultraviolet to millimetre wavelengths. Resolved submillimetre observations constrained the morphology of the parent belt to be a narrow ring at a radius of $\sim80$ au, with a width of about $\sim10$ au \citep{Kennedy2018_HR4796}. In scattered light, this ring also appears very narrow 
\citep{Schneider1999,Thalmann2011,Lagrange2012_HR4796,Wahhaj2014,Rodigas2015,Perrin2015,Milli2015} and is surrounded by a fainter halo extending up to 1000 au \citep{Schneider2018}, likely consisting of small particles more affected by the radiation pressure than the larger particles that remain on orbits within the ring. Analysis of the geometry of the ring at high-angular resolution showed that it has an intrinsic eccentricity of about 7\% \cite[e.g.][]{Milli2017}. The azimuthal brightness distribution is strongly asymmetric \citep{Milli2017}. In the NIR, this system was the first debris disc for which the phase function of the scattering particles could be retrieved over a wide range of scattering angles, showing a very strong peak of forward-scattering compatible with  particles of a few tens of microns in size and a slight backward-scattering behaviour compatible with the presence of aggregates. Near-infrared polarimetric observations confirm this brightness asymmetry \citep{Milli2015,Perrin2015}. 

Polarimetry is a major remote sensing tool for understanding the nature of scattering particles. In this work, our goal is to reveal the surface brightness of the HR\,4796 ring in polarized optical light. We describe our observations in Sect. \ref{sec_obs}, explain our method to extract the polarised phase function in Sect. \ref{sec_method} and discuss the scattering properties of the ring in Sect. \ref{sec_modelling} before concluding in Sect. \ref{sec_conclusions}.


\section{Observations}
\label{sec_obs}

\subsection{Instrumental setup}

The star HR\,4796\,A was observed on the night of 24 May, 2016, with the Spectro-Polarimetric High-contrast Exoplanet REsearch instrument \cite[SPHERE][]{Beuzit2019}. SPHERE is a high-contrast imager fed by an extreme adaptive optics (AO) system \citep{Sauvage2016_SAXO} to correct for the atmospheric turbulence and static aberrations. 
These observations were part of the Guaranteed Time Observations of the SPHERE consortium\footnote{ESO program 097.C-0523(A)} and made use of the visible subsystem Zurich IMaging POLarimeter \cite[ZIMPOL,][]{Schmid2018} to observe the star in linear polarized light. 
The polarimetric mode of ZIMPOL makes use of a special modulating/demodulating CCD synchronised with a fast-switching ferromagnetic liquid crystal retarder in order to record the signal in the two orthogonal linear polarisation directions almost simultaneously through the same pixels of the detectors. This technique is tuned for a very high polarimetric contrast around the star to reveal the circumstellar environment and beat residual noise originating from uncorrected atmospheric speckles  and quasi-static speckles. 
We used the very broad band VBB filter ($\lambda_c=735.4$~nm, $\Delta\lambda=290.5$~nm) to obtain the best sensitivity. ZIMPOL offers two polarimetric modes: fast polarimetry (Fast Pol) with a 1kHz-modulation, high pixel-gain but higher readout noise, and slow polarimetry (Slow Pol) with low gain and low readout noise for longer integrations. We interleaved deep, saturated Slow Pol images to reach the highest sensitivity for the detection of the disc, with shorter Fast Pol images with the neutral density filter ND1 to obtain unsaturated frames of the star. In addition, we used the field-tracking mode of the derotator called P2 to stabilise the field and used five different offset positions of the derotator of $0^\circ$, $30^\circ$, $60^\circ$, $120^\circ$ and $150^\circ$ to provide some additional diversity and reduce the noise in the final combined image. In Slow Pol, we recorded two polarimetric cycles for each offset position. One polarimetric cycle is made of images recorded at four half-wave plate positions to measure the Stokes parameters $+Q$, $-Q$, $+U$ and $-U$ . We measured eight frames (NDIT) of 10~s (DIT) integration at each half-wave plate position, which makes a total on-source exposure time in Slow Pol of 53~min. In Fast Pol, we obtained four polarimetric cycles per derotation offset position, with four frames (NDIT) of 1.2~s (DIT) integration per half-wave plate position, for a total on-source exposure time of 6~min.

Despite very good seeing conditions from 0.4\arcsec{} to 0.7\arcsec{} and fair coherence time from 3 to 6~ms, the ground wind speed was very low, below 2~m/s, and most of the time was below 1~m/s, which affected the quality of the observations. SPHERE indeed suffered from a degradation in the image quality in low-wind conditions, referred to as the low-wind effect \cite[LWE,][]{Sauvage2016,Milli2018}. In these conditions of insufficient air flow in the dome, the air around the cold telescope spider becomes cooler than ambient. This creates disturbances in the wavefront that are barely seen by the instrument wavefront sensor, and the point-spread function (PSF) displays bright side lobes, moving around the central core on a typical timescale of a second. As a result, the full width at half maximum (FWHM) of the long-exposure PSF is not the diffraction limit (19~mas) but reaches about 30~mas and can degrade even to 40~mas for lower-quality exposures. As this phenomenon evolves on timescales much longer than the polarimetric modulation (27~Hz in Slow Pol), the contrast in polarimetry is not affected, but the resolution is lower and the determination of the star centre is less accurate. 

\subsection{Data reduction}

\begin{figure*}
           \centering
   \includegraphics[width=\hsize]{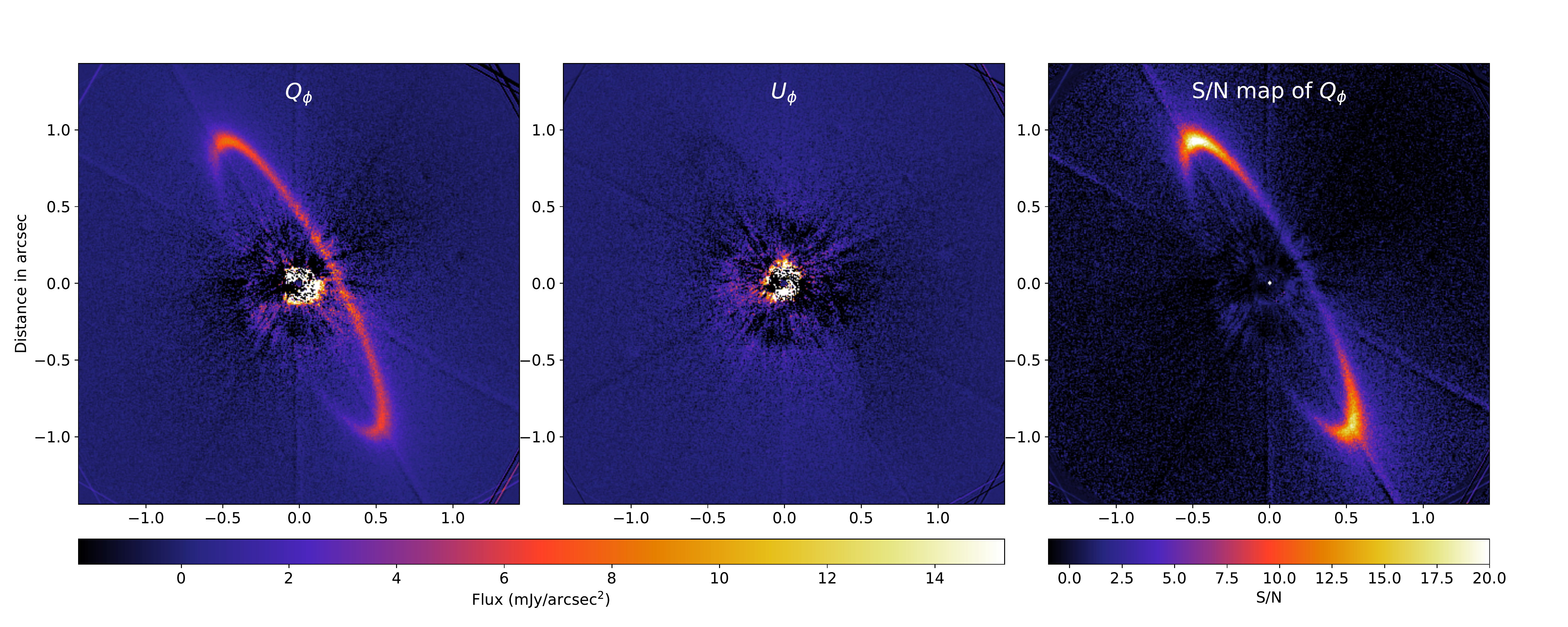}
   \caption{Final images of the azimuthal Stokes $Q_\phi$ (left) and $U_\phi$ (middle), calibrated in mJy/arcsec$^2$. North is up and east to the left. The right image is the S/N map of the $Q_\phi$ signal expressed per pixel.} 
              \label{fig_Qphi_Uphi}
    \end{figure*}

We reduced the data with a custom pipeline to derive the Stokes parameters $I$, $Q,$ and $U$ from each polarimetric cycle. The saturated Slow Pol images were used to detect the disc, while the Fast Pol unsaturated images were used for flux calibration. We initially determined the star centre in the Fast Pol unsaturated images with a Gaussian fit, and used these values to recentre the Slow Pol saturated images. While this yielded sufficient polarimetric attenuation of the stellar signal to detect the disc along the semi-major axis with a high signal-to-noise (S/N) (Olofsson et al. in prep), strong residuals are still present along the semi-minor axis of the disc below 0.4\arcsec\  due to the unusually large tip-tilt jitter induced by the LWE. After several tests, we found that using the barycentre of the saturated pixels to recentre the Slow Pol images provided a much higher stellar attenuation and correction of the beamshift \cite[see][]{Schmid2018} to reveal the disc semi-minor axis. We estimate the accuracy of the recentring to be better than one pixel, that is, 7.2~mas along the detector vertical axis and 3.6~mas along the detector horizontal axis (1.3\% of the ring semi-minor axis). 
For each polarimetric cycle, we derived the Stokes $I$, $Q,$ and $U$ following the steps outlined in \citet{Engler2017}, and corrected the instrumental polarisation (IP) by subtracting from Q and U a scaled version of $I$. 
The residual telescope polarisation was found to be between 0.2 and 0.4\% with an average of about 0.3\%, assuming the central star is not polarized, which is consistent with the average residual telescope polarisation of $\sim0.4\%$ derived for the VBB filter by \citet{Schmid2018}.  As we expect the disc polarisation signal to be purely tangential or radial in the case of single-scattering by an optically thin disc, we use the azimuthal Stokes parameter $Q_\phi$ and $U_\phi$ defined as  $Q_\phi=Q \text{cos}2\phi+U\text{sin}2\phi$ and $U_\phi=Q \text{sin}2\phi-U\text{cos}2\phi$ \citep{Schmid2006}, where $\phi$ is the polar angle between north and the point of interest measured from the north over east (the position angle). 
$Q_\phi>0$ is equivalent to a tangential polarisation component while $Q_\phi<0$ indicates a radial polarisation. The component $U_\phi$ describes the polarisation in the directions $\pm45^\circ$ with respect to the radial direction. We then combined all the cycles together after re-aligning the images with the north up and rebinned the original pixels along the horizontal axis to get square pixels of 7.2~mas in  size. The $Q_\phi$ and $U_\phi$ are shown in Fig. \ref{fig_Qphi_Uphi}, after conversion to milliJanskies per arcsecond squared. For the conversion, we estimated the star flux as the total flux contained in a circular aperture of radius 100~px (720~mas) of the mean unsaturated Fast Pol PSF, corrected by the difference between the Fast Pol and Slow Pol setup (different detector gain and integration time), and we used a stellar flux density of $16.1 \pm 0.1$~Jy and a pixel surface area of $7.2\times7.2$ square milliarcseconds. The flux density of $16.1 \pm 0.1$~Jy was obtained after converting the V magnitude of $5.774 \pm0.009$ \citep{Hog2000} to Janskies and correcting for the different filter response between the Tycho V and Zimpol VBB using a blackbody spectrum at 9730~K typical of an A0V star.

\section{Analysis and extraction of the phase function}
\label{sec_method}

\subsection{Morphology and surface brightness}
\label{sec_morpho}

As shown on the left of Fig. \ref{fig_Qphi_Uphi}, the disc is detected as a positive signal in the Stokes $Q_\phi$ image, indicating tangential polarisation, and it has an elliptical shape. The west side of the ellipse is bright and well detected while the east side is fainter and even lost among the residual noise of the image below a radius of $\sim$0.5\arcsec.  The S/N (Fig. \ref{fig_Qphi_Uphi} right) is highest in the north ansa where it reaches 23, and decreases on the west side to reach a value of 3 at the semi-minor axis of the disc. This S/N is expressed here per pixel (7.2~mas), but can be up to four times higher expressed per resolution element (FWHM of 30~mas) if the noise is Gaussian. In the background, five faint radial lines are only just visible and correspond to the vertical line of pixels passing through the star, for the five position angles of the detector. Under heavy saturation, this pixel line is indeed brighter due to CCD frame transfer smearing, and is not entirely removed by the polarimetric subtraction; this does not affect the analysis however. 
The $U_\phi$ image (Fig. \ref{fig_Qphi_Uphi} middle) shows the same noise structure with bright residuals close to the star below 0.2\arcsec{} but no disc emission, as expected from single-scattering. A very faint negative shadow of the ring is detectable with a minimum flux of -0.05 mJy/arcsec$^2$. We do not believe this corresponds to a true disc signal in $U_\phi$ but attribute it to the effect of the convolution of the astrophysical signal with the PSF, which happens even if $U_\phi$ is zero before convolution, as shown in Appendix A of \citet{Engler2018}.

As already noticed in the optical and NIR, the disc ansae are asymmetric between the NE and SW. 
The maximum surface brightness for the polarised intensity $Q_\phi$ of the NE ansa is $8.3 \pm 0.7$ mJy/arcsec$^2$ while the SW ansa reaches only $6.2 \pm 0.6$ mJy/arcsec$^2$, that is, the SW/NE asymmetry factor is $0.75 \pm 0.1$, in agreement with \citet{Schneider2009} who measured a value of $0.74 \pm 0.07$ in broadband HST/STIS images ($\lambda_c=575$ nm). Because HST/STIS only measures Stokes I, this already indicates that the polarisation fraction at the two ansae is the same and is symmetric. 

\begin{figure}
   \centering
   \includegraphics[width=\hsize]{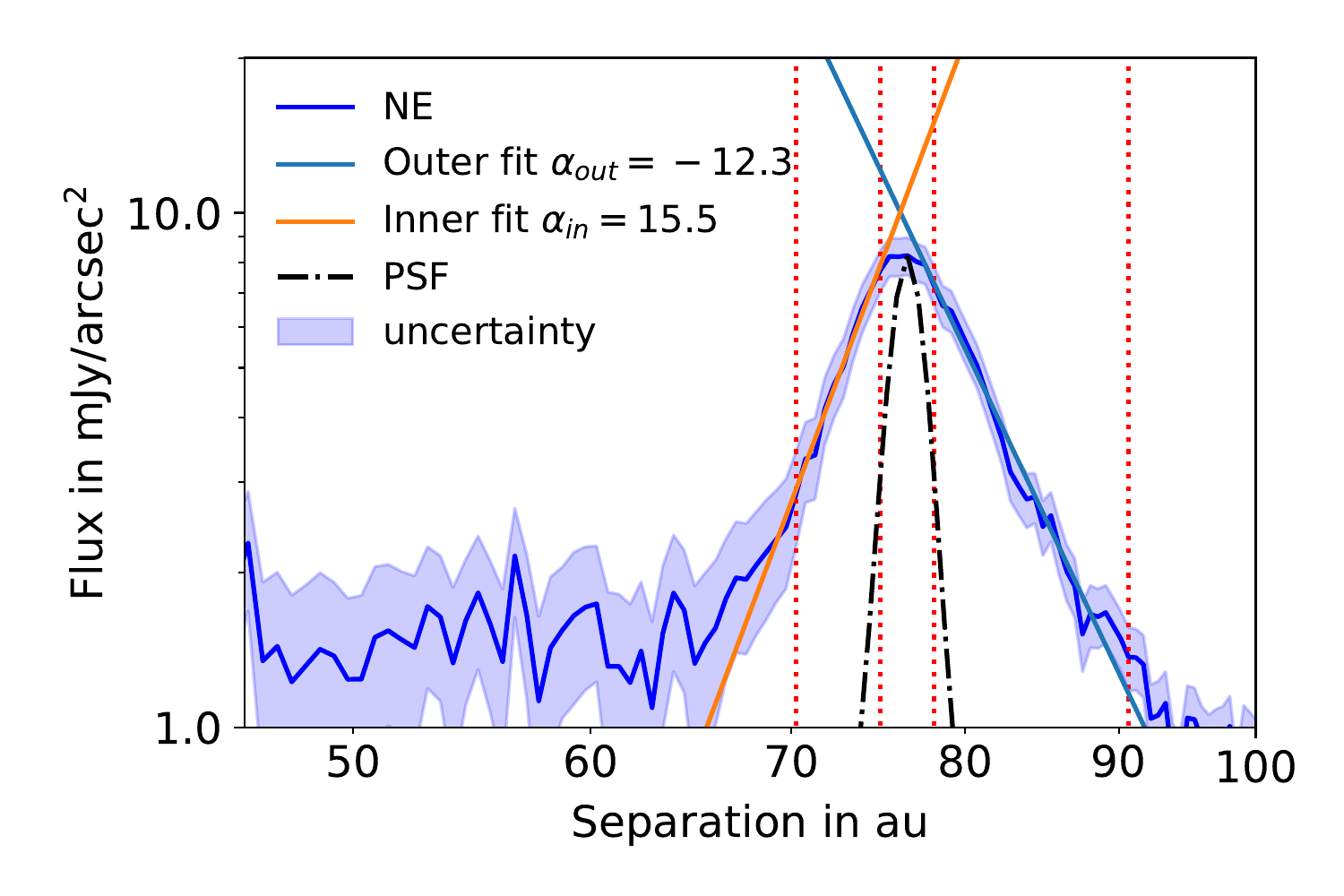}
   \includegraphics[width=\hsize]{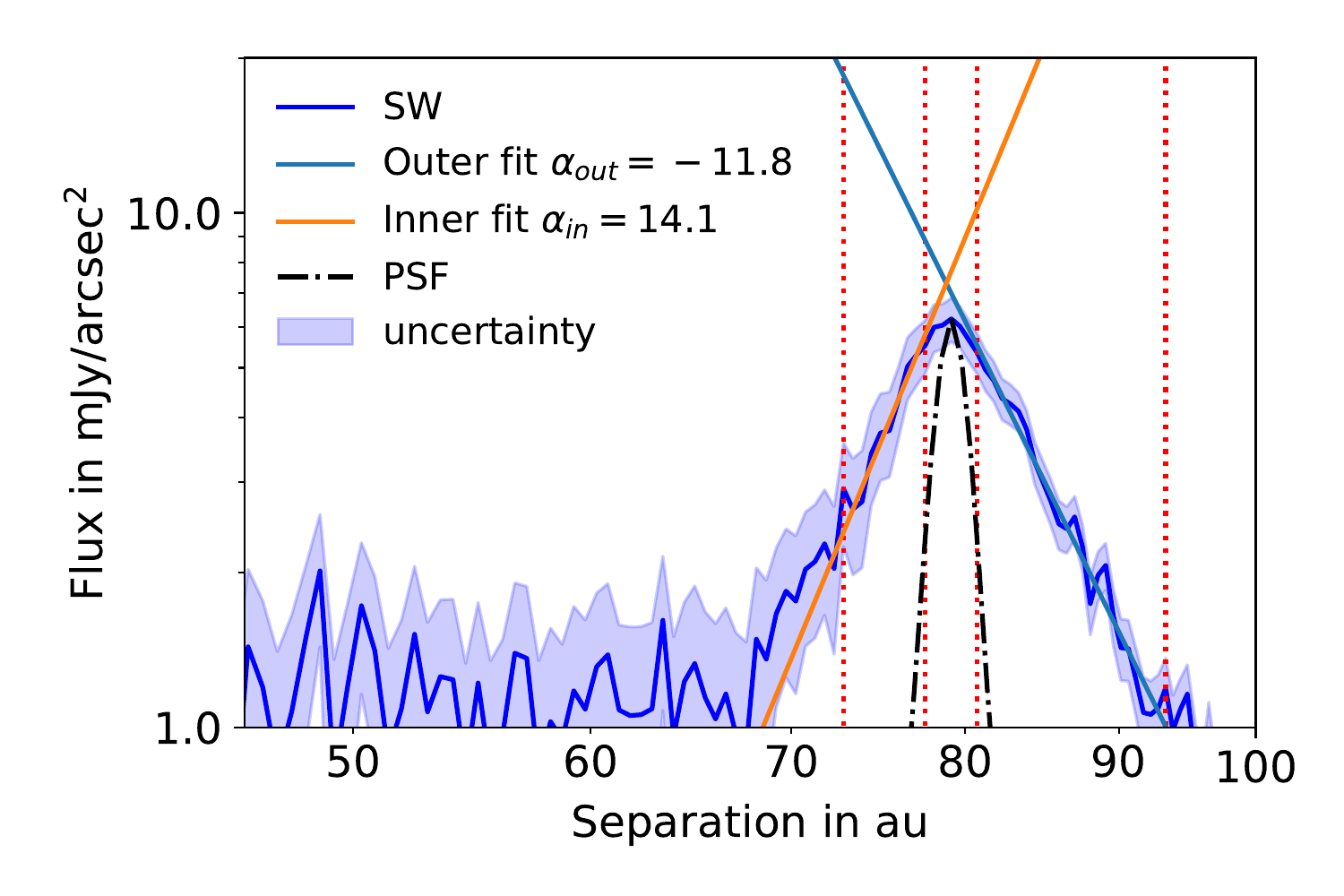}
   \caption{Radial profile along the NE (top) and SW (bottom) ansae of the disc. The vertical red dotted lines show the boundaries used for the fit of a power law to the inner and outer profile. The profile of the PSF (in black) is indicated as a reference.}
    \label{fig_NE_ansa}%
    \end{figure}

The disc is clearly resolved radially. We show in Fig. \ref{fig_NE_ansa} the radial profiles along the major axis, in the NE and SW, respectively, where the blue shaded area represents the $1\sigma$ noise measured radially in the $Q_\phi$ image. The black line shows the average PSF profile centred at the peak brightness of the radial profile. The maximum brightness occurs at 1.087\arcsec and 1.051\arcsec , respectively. The FWHM of the ring is 136 mas $\pm$ 33mas and 144 mas $\pm$ 41 mas for the NE and SW sides, respectively. This is larger than what was typically measured at NIR wavelengths: 111 mas $\pm$ 43mas and 137 mas $\pm$ 50mas in the H band \citep{Milli2017} but smaller than optical measurements at bluer wavelengths \cite[184mas $\pm$ 10mas, ][]{Schneider2009}. 


The steepness of the inner and outer profiles is consistent within error bars with that measured in the NIR. This is important confirmation, because the star-suppression algorithm used for NIR unpolarised light \cite[Angular Differential Images, ADI][]{Marois2006} can bias the measurement \citep{Milli2012}, especially for the inner slopes, and non-ADI measurements have large error bars. We fitted a power law to the inner and outer profile, as already done in \citet{Milli2017} and show the result of the fit in Fig.  \ref{fig_NE_ansa}  (coloured lines).  We derived values of 15.5 $\pm$ 1.6 and 14.1 $\pm$ 2.6 for the inner slopes $\alpha_\text{in}$ of the NE and SW ansae, respectively. The outer slopes $\alpha_\text{out}$ are -12.3 $\pm$ 0.9 and  -11.8 $\pm$ 0.9  for the NE and SW ansae, respectively. This is slightly shallower than the NIR profiles, for which the mean slopes as measured in the non-ADI images were -13.5 $\pm$ 3.1 and -12.2 $\pm$ 1.9, although the difference is not significant.

 
We derived the morphology of the disc following the approach detailed in \citet{Milli2017}. We first regularly sampled the elliptical ring, with one point every resolution element. To do so, we extracted radial profiles passing through the star and crossing the ring and we determined the location of the maximum brightness of the disc by fitting a two-component power law. The resulting data points are shown in Appendix \ref{App_MCMC} (inset of Fig. \ref{fig_corner_plot_MCMC}). As the ring is not detected everywhere,  we selected only those radial profiles with at least one pixel with a S/N greater than three, corresponding to profiles with a position angle between $\sim50^\circ$ and $\sim90^\circ$. We used a Markov Chain Monte Carlo algorithm \cite[MCMC, using the affine-invariant Python implementation emcee;][]{Foreman-Mackey2013} to derive the best ellipse passing through those points using the approach described in \citet{Ray2008}. We first derived the projected parameter of the ellipse in the plane of the sky:  the projected semi-major axis $a^\prime$, the projected semi-minor axis $b^\prime$, the offsets in right ascension and declination of the ellipse centre with respect to the star location $\Delta\alpha$ and $\Delta\delta$, and the position angle $PA$. These parameters are given in Table \ref{tab_ring_morphological} in the rows corresponding to "projected ellipse", together with the uncertainty measured directly on the posterior probability density function of the fitted parameters. 

\begin{figure}
           \centering
   \includegraphics[width=\hsize]{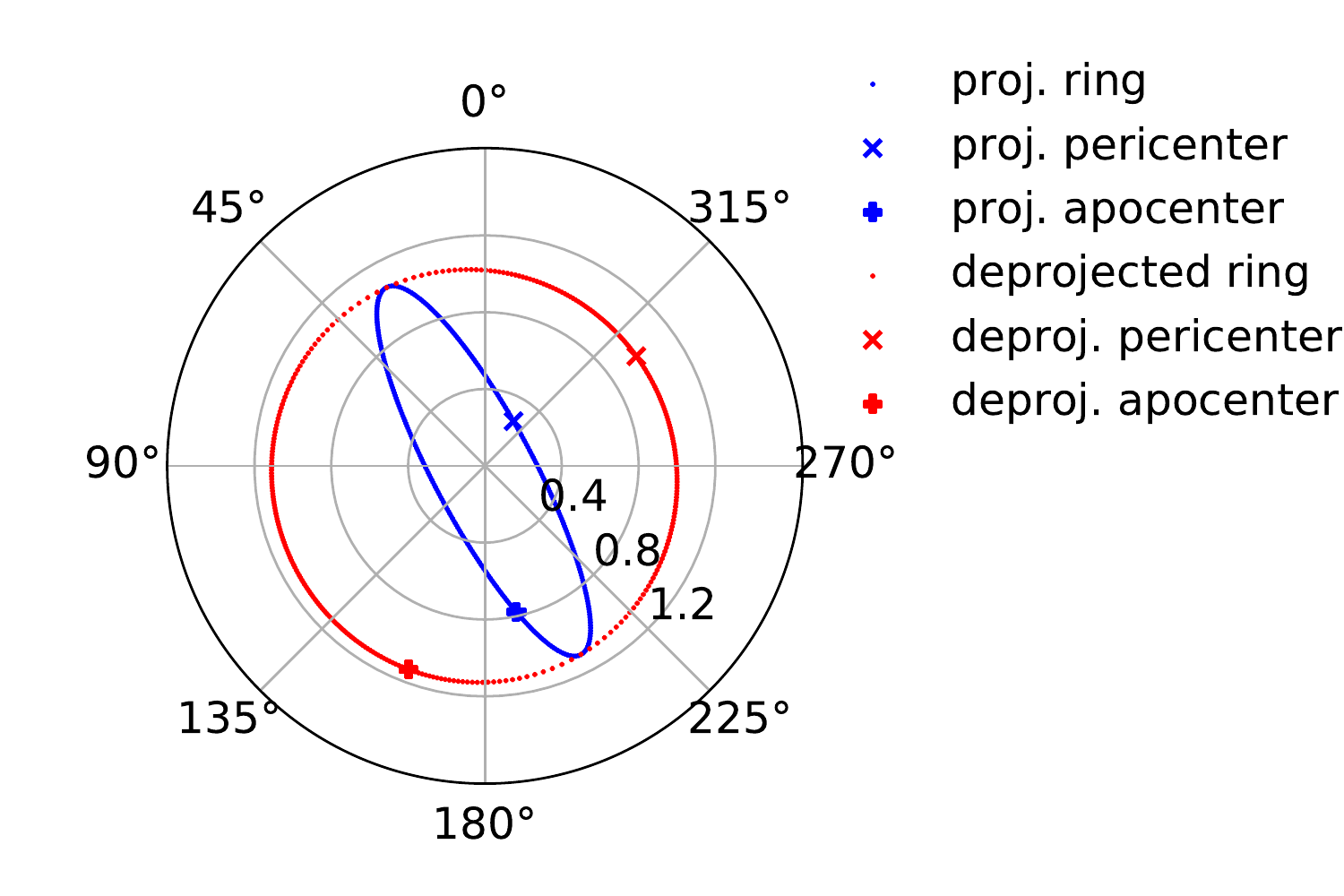}
   \caption{Best-fit ellipse and location of the pericentre/apocentre for the projected (blue) and deprojected (red) ring. Radial graduations are in arcseconds.} 
              \label{fig_ellipse_polar}%
    \end{figure}

Using the Kowalsky deprojection technique \citep{Smart1930}, we derived the parameters of the true ellipse described by the dust particles in the orbital plane: the true semi-major axis $a$, the eccentricity $e$, the inclination $i$, the argument of pericentre $\omega,$ and the longitude of the ascending node $\Omega$. The result is given in Table \ref{tab_ring_morphological}, in the rows corresponding to "deprojected ellipse".  We show the corresponding projected and deprojected ellipses in a polar plot in Fig. \ref{fig_ellipse_polar}, along with the position of the pericentre. These measurements are compatible (at $3\sigma$) with those already published in \citet{Milli2017} using the IRDIS sub-system in the H band (Table \ref{tab_ring_morphological}, right column), a different image processing technique (ADI), and an identical measurement procedure. 

\begin{table}
\caption{Projected and deprojected ring parameters. The error is given at a $3\sigma$ level and contains only the statistical error from the fit and no systematic error from the true north or star registration.}
\label{tab_ring_morphological}
\centering
\begin{tabular}{c c |c c c}
& & & this work & IRDIS H \citep{Milli2017}\\
\hline 
\parbox[t]{2mm}{\multirow{5}{*}{\rotatebox[origin=c]{90}{Projected}}} & \parbox[t]{2mm}{\multirow{5}{*}{\rotatebox[origin=c]{90}{ellipse}}} &                                                         $a^\prime$(mas) & $ 1073 \pm     4$ &  $ 1064 \pm     6$ \\
                                                 & &   $b^\prime$(mas) & $  260 \pm     7$ &  $  252 \pm     4$  \\
                                                 & &   $\Delta\alpha$(mas) & $   -6 \pm     4$  & $   -4 \pm     4$  \\
                                                 & &   $\Delta\delta$(mas) & $  -27 \pm     4$  & $  -28 \pm     5$ \\
                                                 & &   PA($^\circ$) &  $27.9 \pm  0.2$  & $27.69 \pm  0.26$ \\
\hline
\parbox[t]{2mm}{\multirow{5}{*}{\rotatebox[origin=c]{90}{Deprojected}}} & \parbox[t]{2mm}{\multirow{5}{*}{\rotatebox[origin=c]{90}{ellipse}}}                                                 & $a$(mas) & $ 1076 \pm     6$  & $ 1066 \pm     6$   \\
                                                   &   & $e$ & $0.072 \pm 0.037$ & $0.070 \pm 0.011$  \\
                                                    &  & $i$($^\circ$)\tablefootmark{a}  & $76.0 \pm  1.2$ & $76.33 \pm  0.24$ \\
                                                     & & $\omega$($^\circ$) &  $-74.2 \pm 11.9$ & $-72.44\pm  5.10$\\
                                                      & & $\Omega$($^\circ$)\tablefootmark{a} &  $27.9 \pm  0.6$  & $27.71 \pm  0.25$ \\
\hline
\end{tabular}
\tablefoot{
\tablefoottext{a}{We followed the previous conventions used for this system with an ascending node $\sim 28^\circ$ (measured from north, anti-clockwise) for an inclination of $\sim76^\circ$ ($0^\circ$ means pole-on) but as noted in \citet{Kennedy2018_HR4796}, the west side is closer to Earth, and therefore the inclination should be strictly  $\sim104^\circ$ (or the node should be $\sim208^\circ$ and the inclination retained).}
}
\end{table}

\subsection{Polarised phase function}
\label{sec_SPF_extraction}

\begin{figure}
   \centering
   \includegraphics[width=\hsize]{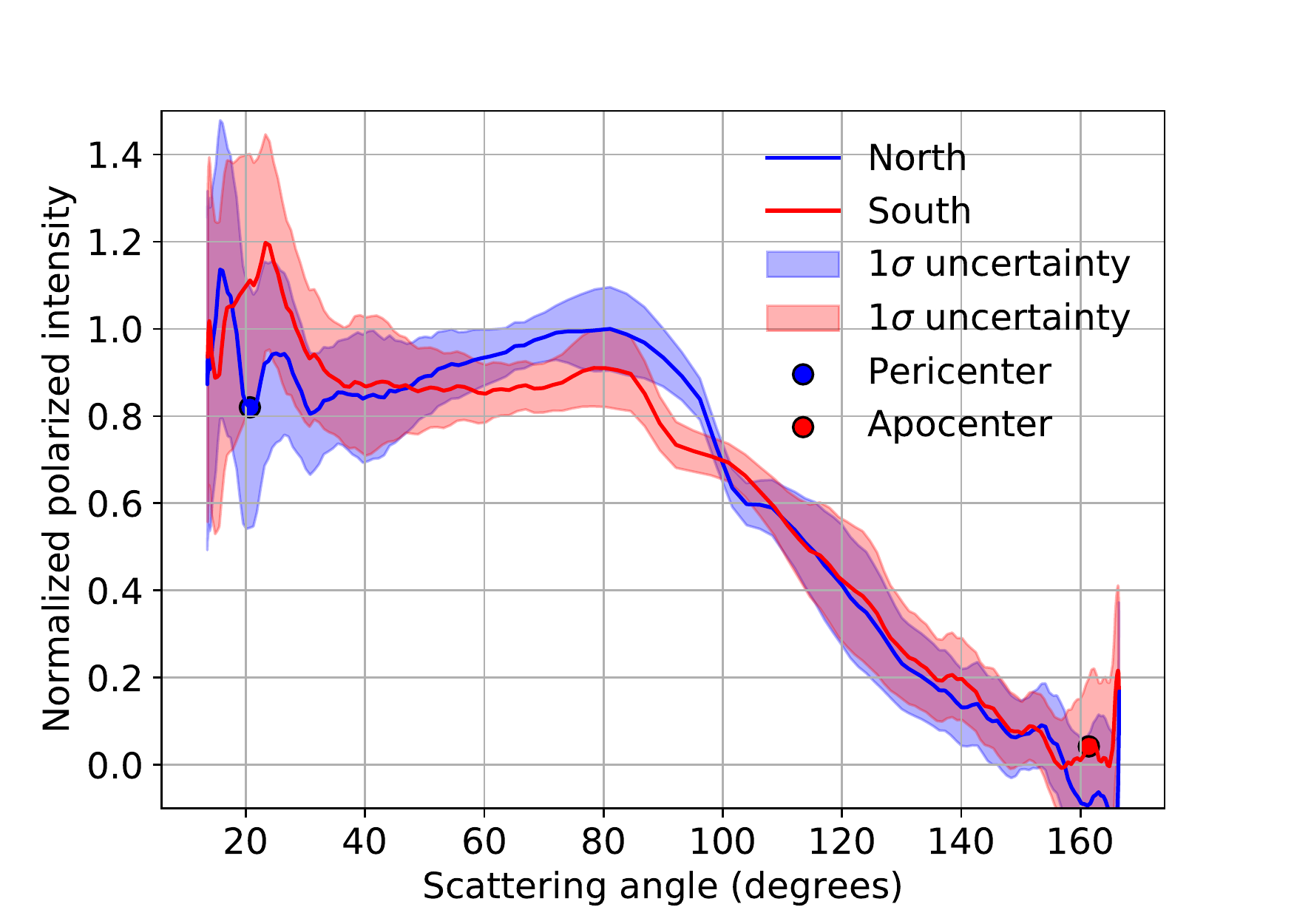}
   \caption{Polarized SPF as extracted on the north and south side of the disc. The curves were normalised to one at $\varphi=80^\circ$ on the north side.}
    \label{fig_pol_phase_functioni}%
    \end{figure}

We used the accurate morphology of the disc derived in Sect. \ref{sec_morpho} to extract the polarised scattering phase function (pSPF) of the dust in polarised light. Several assumptions are required to do so: one must assume a flat disc (small vertical extension above the mid-plane compared to the image resolution, that is, scale height smaller than $\sim2.5$\,au  equivalent to a disc aspect ratio smaller than 3\%), so that each point of the ring can be associated with a unique value of the scattering angle. One must also assume that the dust density is azimuthally uniform and that the dust properties do not change with azimuth. The  ALMA dust continuum image at 880\micron{}  \citep{Kennedy2018_HR4796} does not show any significant asymmetry besides a pair of $3\sigma$ blobs on the east side. This assumption is therefore reasonable for millimetre-sized particles. Moreover, we showed in Sect. \ref{sec_morpho} that the polarised fraction of the dust is identical between the two ansae of the disc, which also supports this assumption for micron-sized particles. We also considered the west side as the forward-scattering side of the disc, as shown in \citet{Milli2017}.
We then followed the methodology detailed in \citet{Milli2017} to derive the scattering  angle $\varphi$ associated to any point of the ring at a position angle $\theta$ in the plane of the sky: 

\begin{equation}
\label{eq_phi}
\varphi = \arcsin\left( \frac{1}{\sqrt{\sin^2(\theta-\Omega)/\cos^2i+\cos^2(\theta-\Omega)}} \right)
.\end{equation}
We placed elliptical apertures along the ring, with the major axis perpendicular to the PA of the ring, a length equal to 137~mas (FWHM of the ring along the NE ansa) and a major to minor axis ratio dictated by the ring inclination $\sec i = 4.135$. These elliptical apertures account for the projection effect of the disc: this technique is equivalent to using circular apertures around the circumference of a face-on ring. We found that using apertures with a major axis equal to the
disc FWHM at the ansae maximised the S/N in the aperture photometry. The extracted photometry needs to be corrected by the illumination of the central star and by the effect of the convolution by the PSF. The first correction factor accounts for the fact that the starlight received and scattered by the dust particles depends on the inverse of the squared deprojected distance from the star.
The second correction factor accounts for the dilution of the flux due to the size of the PSF. Because of the geometry of the disc and the use of elliptical apertures, the convolution has a different effect along the ring and this needs to be corrected for. This is described in Appendix \ref{App_convolution}. The result after taking those two correction factors into account and after normalisation to unity at $80^\circ$ for the north ansa is shown in Fig. \ref{fig_pol_phase_functioni}. We measured the pSPF on the north and south ansa independently. Due to the inclination of the system, we can probe scattering angles from $13^\circ$ (closest part of the ring on the W side) to $167^\circ$ (E side, without detectable polarised emission). 

The uncertainties presented in Fig. \ref{fig_pol_phase_functioni} take into account three sources. The measurement uncertainty from the aperture photometry is the largest source of error, especially at short separations from the star. We also included the error on the illumination factor, stemming from the uncertainty on the disc geometry and the error on the convolution correction factor. To derive the latter term, we propagated the uncertainty on the pSPF from the model described in Appendix \ref{App_convolution} down to the impact of the convolution by the PSF. This dominates the error for scattering angles between $110^\circ$ and $140^\circ$. 

The curve shows several interesting features. The polarised emission from the disc is not detected for scattering angles above $\sim150^\circ$ on either side. In particular we do not have the sensitivity to detect whether the polarised fraction changes sign at large scattering angles, as is commonly seen for comets. With increased sensitivity we would have seen a negative signal in $Q_\phi$. Our measurement would be compatible with an inversion around $160^\circ$. Interestingly $160^\circ$ is also the typical value for comets in the solar system \citep{Kiselev2015}. 

The pSPF is identical for both the N and S sides on the backward scattering side beyond $100^\circ$ with a similar slope. There is a local maximum of the pSPF at $83^\circ$ for both sides of the disc. The global maximum between $13^\circ$ and  $167^\circ$ is reached at smaller scattering angles of $15^\circ$ and $25^\circ$  for the N and S sides, respectively. The noise is higher in this region: while the oscillations seen below $\sim40^\circ$ are likely artefacts from the residual stellar light, the overall trend showing an inflection in the pSPF at $\sim40^\circ$ seen in both sides of the disc and a change of sign of the first derivative of the pSPF is real. 

Between scattering angles of $60^\circ$ and $100^\circ$, we see a difference in pSPF between the two sides of the disc. The illumination factor due to the eccentricity of the disc is not enough to explain the asymmetry, as already reported in past studies \cite[e.g.][]{Wahhaj2014}: the SW/NE illumination factor is 0.91 in the ansae whereas we measured a brightness ratio of $0.75 \pm 0.1$. Either the dust density is asymmetric between the NE and SW sides or the dust properties are different. We note also that the slope of the pSPF on the forward scattering side is also different between the north and south sides, which again might support either a difference in dust density, dust properties, or both. 

This brightness asymmetry between the two sides of the disc was initially revealed from low-resolution mid-infrared (MIR) imaging \citep{Wyatt1999,Telesco2000,Moerchen2011} and is attributed to pericentre glow \citep{Pan2016}. Both the higher luminosity of the  NE ansa and the 10K colour difference at MIR wavelengths can be explained by the eccentricity of the disc. However, for the eccentricity to remain small ($e<0.1$), \citet{Moerchen2011} show that the pericentre has to stay close to the NE ansa. This contradicts the high-angular resolution images of the disc which consistently showed that the argument of pericentre is closer to the semi-minor axis of the disc on the NW side than to the NE ansa \cite[][and also this study]{Schneider2009,Thalmann2011,Wahhaj2014,Rodigas2015,Milli2017}. The MIR studies did however not consider the dependence of the collision rate with azimuth. In an eccentric disc, the Keplerian orbital velocity is larger at pericentre than at apocentre, causing parent bodies to spend more time near the apocentre, but collisions happen more frequently near the pericentre. Olofsson et al. (in prep) developed an analytical model accounting for this enhanced collision rate near pericentre and applied it to fit the ZIMPOL observations of the ansae of HR\,4796. These latter authors show that the brightness ratio between the N and S ansae can be explained with the disc eccentricity and location of the pericentre measured in this study if small dust particles are preferentially released near the pericentre. They also show that their model is compatible with the MIR observations presented in  \citet{Moerchen2011} and the submillimetre observations presented in \citet{Kennedy2018_HR4796} that do not show any asymmetry. To mitigate this effect and still be able to interpret the pSPF and discuss the dust properties (section \ref{sec_modelling}), we averaged the pSPF between the N and S sides. We note that this simplification does not affect the conclusions that can be drawn from this study regarding the properties of the dust particles. 


\subsection{Polarised fraction}
\label{sec_polar_frac}

Because this ZIMPOL data set was obtained in field tracking mode, it is not possible to extract the intensity image of the disc and derive the polarised fraction over the ring. However, we can compare our measured polarised surface brightness (SB) with the value reported in unpolarised light in the optical. \citet{Rodigas2014} reports an average SB of  29.0~mJy/arcsec$^2$ in the ansae at the VBB central wavelength with HST, as interpolated from their Fig.~6 using a stellar flux of 16.1~mJy.  
Compared to the averaged polarised SB measured in the ansae in this study, this implies a polarised fraction of 40\% $\pm$ 26\%. 
This result includes a correction factor to take into account the effect of convolution. Using our best disc model described in Appendix \ref{App_convolution}, we found that the convolution attenuates the flux in the ansae by a factor 1.6. The error bar includes the error on the polarised SB, an ad hoc 10\% uncertainty on the unpolarised SB from \citet{Rodigas2014} interpolated linearly at 735~nm, and 5\% uncertainty on the correction factor for the convolution as observed by testing different PSF.

\citet{Perrin2015} analysed the polarised scattered light of the ring in the Ks band, and derived the polarisation fraction from $40^\circ$ to $110^\circ$. The polarised fraction at $90^\circ$ in this band is $\sim$25\%, which is smaller than the $40\% \pm 26\%$ measured here in the optical, although still compatible within error bars. 

\section{Modelling and discussion}
\label{sec_modelling}

\subsection{Compatibility with the Henyey-Greenstein approximation} 

A common and simple way to describe the polarised phase function of debris discs is to use the Henyey-Greenstein analytical prescription of the SPF \cite[HG,][]{Henyey1941}, parametrised by the anisotropic scattering factor g (between -1 and 1), combined with the Rayleigh scattering polarisation fraction $p$, such that SPF $\times p \propto f(g,\varphi)$ with
\begin{equation}
f(g,\varphi) = \frac{1-g^2}{\left( 1-2g\cos{\varphi} +g^2 \right)^{3/2}}\times\frac{1-\cos^2{\varphi}}{1+\cos^2{\varphi}}
\label{eq_HG}
.\end{equation}
The HG $g$ parameter characterises the shape of the phase function. For isotropic scattering $g=0$, forward scattering particles have $0<g\leq1$ while backward scattering partices have $-1\leq g<1$.
This approach combined to a polarisation description with the same scattering angle dependence as Rayleigh scattering was used for instance for the modelling of HIP~79977 and HD~172555 \citep{Engler2017,Engler2018}. 

We therefore investigated how accurate this representation is for the dust particles surrounding HR\,4796. We are aware that we are likely not in the Rayleigh regime because the particle size $s$ is about 20\micron{} \citep{Milli2017} while the wavelength of observations is $\lambda=735$~nm, which makes the size parameter $x=\frac{2\pi s}{\lambda}\sim170\gg1$.

The best fit is obtained for a value of $g=0.43$, and is shown by the red dashed line in Fig. \ref{fig_fit_MCFOST}. It accurately represents the scattering behaviour beyond $80^\circ$ but fails to accurately capture the behaviour below $80^\circ$, in particular the inflection around $30^\circ$. We estimated the goodness of fit with the measure of the reduced $\chi^2$ and found a minimum value of 0.8. A value of $g=0.43$ probably underestimates the peak of forward scattering of the unpolarised SPF. A similar conclusion was already reached in the NIR, where a HG SPF could not reproduce the behaviour of the SPF in unpolarised light. However, a two-component HG yielded a good fit to the data, with an extremely forward scattering component $g_1=0.99^{+0.01}_{-0.38}$ of weight 83\%, and a slightly backward-scattering second component $g_2=-0.14 \pm 0.006$ of weight 17\% \citep{Milli2017}. Inspired by this fit, we performed a similar fit of a two-component HG combined with the Rayleigh scattering polarisation fraction, proportional to  $w f(g_1,\varphi) + (1-w) f(g_2,\varphi)$. The components $g_1$ and $g_2$ are the HG coefficients of the two components, $w$ and $1-w$ are the corresponding weights between 0 and 1, and $f$ is the function defined in Eq. \ref{eq_HG}. A very good fit ($\chi^2=0.1$) was obtained with a first component strongly forward scattering with  $g_1=0.83^{+0.17}_{-0.30}$ of weight $38\% \pm 33\%$, and a relatively isotropic second component with $g_2=0.09 \pm0.4$ of weight $62\% \pm 30\%$. Despite the large error bars, this model is compatible with the unpolarised SPF derived in the H band.

\subsection{Compatibility with the Mie and DHS theory}
\label{sec_Mie}

In \citet{Milli2015}, we computed the theoretical SPF and polarised fraction for a sample of 7800 dust compositions and sizes, using the Mie theory and the distribution of hollow spheres \cite[DHS,][]{Min2005} as provided in the radiative transfer code MCFOST \citep{Pinte2006}. Here, we reused these models to investigate the compatibility with the new ZIMPOL data, and recap briefly the underlying assumptions. These models are based on a porous dust particle composed of a mixture of astronomical amorphous silicates, carbonaceous refractory material, and water ice partially filling the holes created by porosity. The composition is parametrised by the porosity without ice $P$, a fraction of vacuum removed by the ice $p_\text{H2O}$, and a silicate over organic refractory volume fraction $q_\text{Sior}$. The size of the smallest particles is written $s_\text{min}$. The best polarised SPFs matching the data are those with the minimum particle size 0.1\micron{} using the Mie theory. The best fit is shown in  Fig. \ref{fig_fit_MCFOST} in blue, and more details on the properties of this model can be found in Appendix \ref{App_best_model}. The best fit does not accurately reproduce the measured pSPF because it shows a maximum at $\varphi=55^\circ$ instead of the flat plateau between $30^\circ$ and $80^\circ$. In addition, some resonant oscillations are present in this model below $40^\circ$ because of the spherical geometry of the particles, but they are not observed in the data. 
The polarised fraction of this model is compatible with that measured in HR\,4796 (Fig. \ref{fig_fit_MCFOST_polar_frac}), but the large error bar on this measurement is not very constraining. 
The presence of such small particles in the system is unlikely for several reasons. Firstly, the reflected spectrum of a dust population dominated by sub-micron particles is blue and incompatible with the red spectrum of the disc as derived in \citet{Rodigas2014}. Secondly, these small particles would behave like spherical particles in the Rayleigh scattering regime and we do not see the oscillations typical of this regime in the pSPF. 
Lastly, such small particles would also be blown out of the system by radiation pressure on a short timescale. \citet{Augereau1999}  indeed derived a blowout size of 10 \micron{} from the spectral energy distribution (SED) modelling assuming Mie porous spherical particles of similar composition with the Bruggeman mixing rule. \citet{Arnold2019} showed that considering irregular aggregates for this system increases the blowout size even further, whatever the composition of the particles, and this adds to this inconsistency. These considerations agree with the $\sim20$ \micron{} minimum particle size derived from the SPF analysis in the NIR \citep{Milli2017}.
\citet{Thebault2019} argue that a dynamically active and bright debris disc with more than $10^{-3}$ IR excess can maintain in steady state a population of small particles below the blowout size. However, in the case of HR\,4796\,A, the modelling done in \citet{Milli2015} shows that sub-micron particles cannot contribute to a significant amount of scattered light flux in the optical in order to be compatible with the very red colour of the spectrum of the dust. 

 \begin{figure}
   \centering
   \includegraphics[width=\hsize]{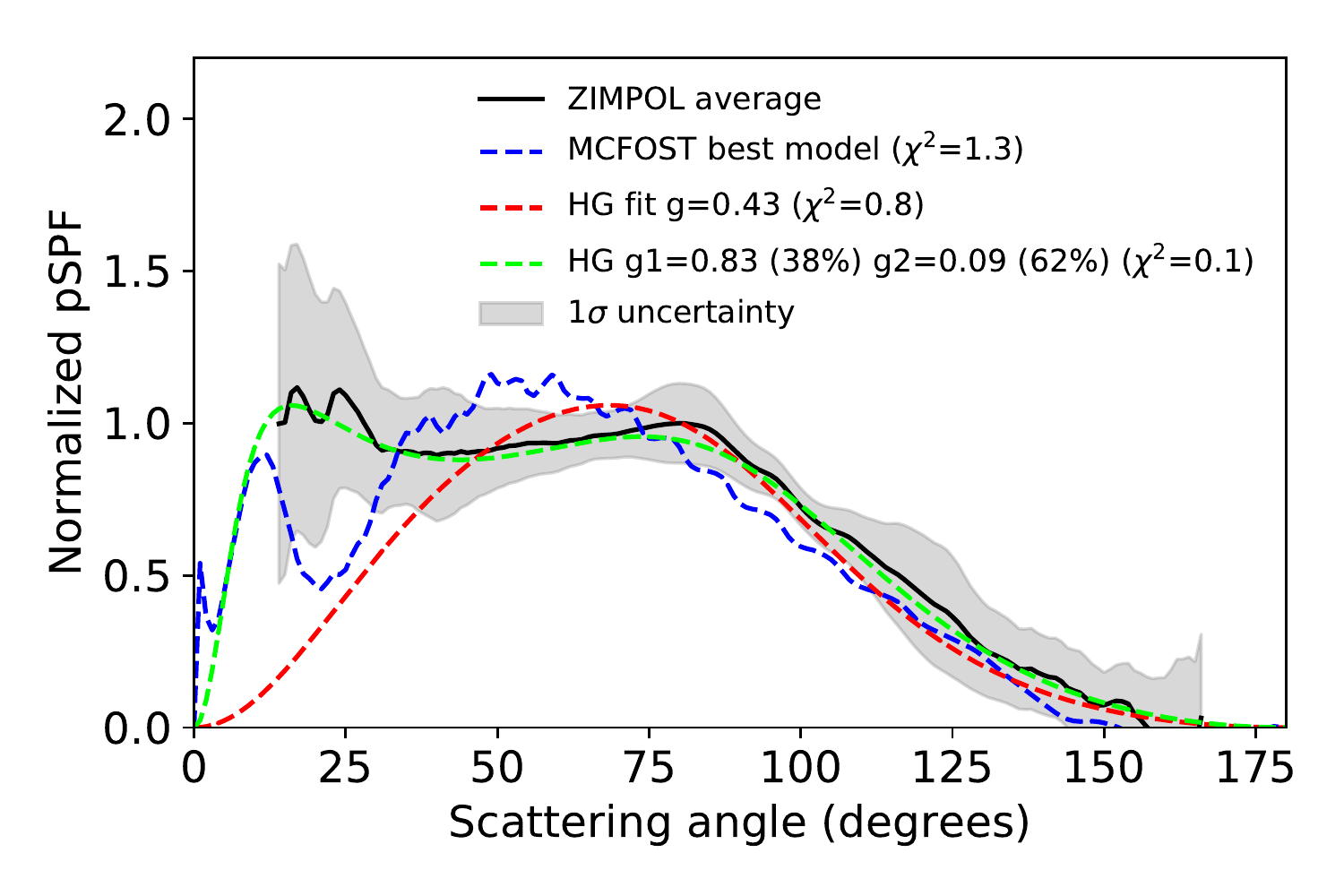}
   \caption{Average polarized phase function (in black) along with the best MCFOST model (in blue) and the best fit of a Heyney Greenstein SPF + Rayleigh polarisation fraction (in red and green).}
    \label{fig_fit_MCFOST}
    \end{figure}

 \begin{figure}
   \centering
   \includegraphics[width=\hsize]{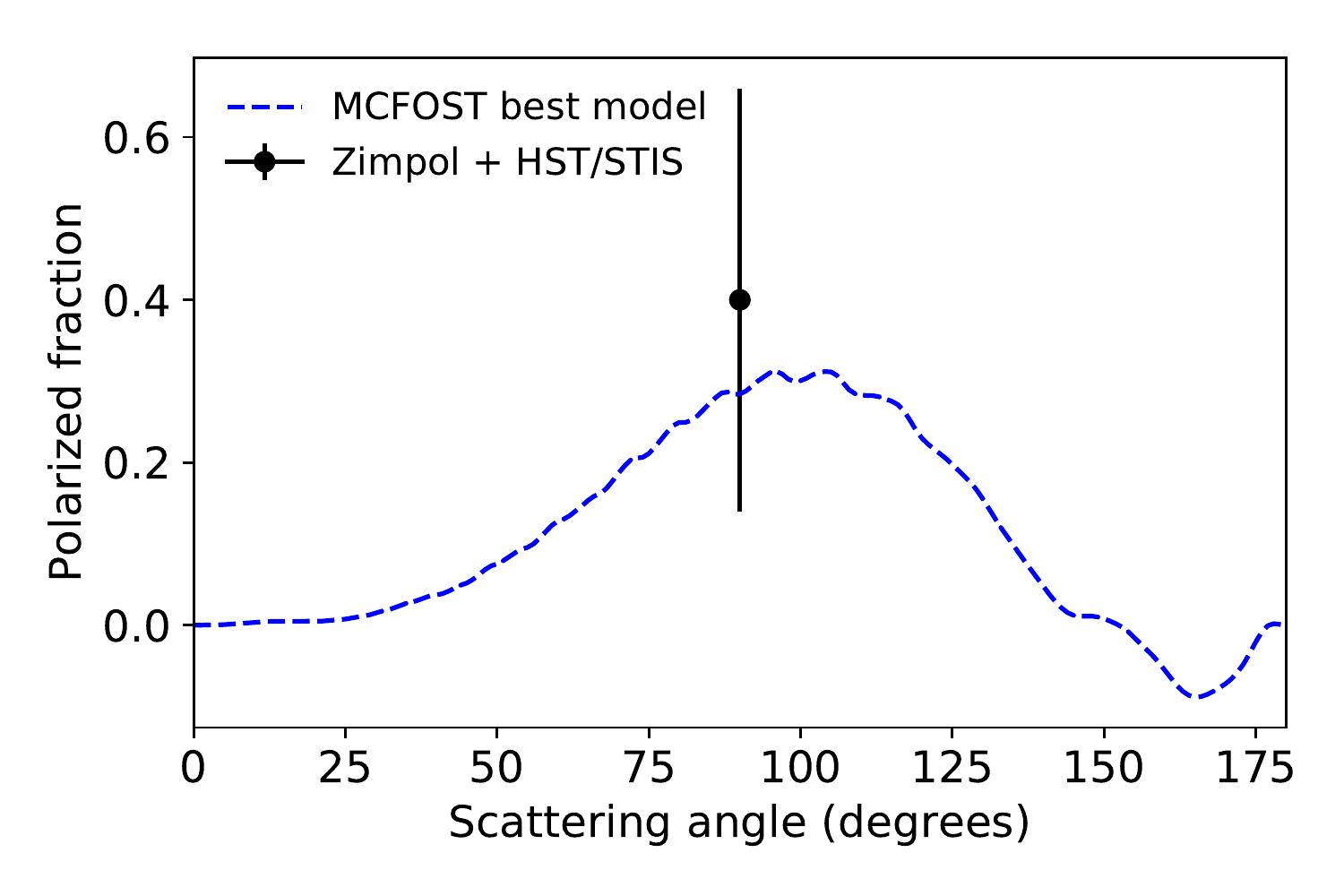}
   \caption{Polarisation fraction corresponding to the MCFOST model shown in Fig. \ref{fig_fit_MCFOST}. The relative error in the polarised fraction at $90^\circ$ is -28.9\%.}
    \label{fig_fit_MCFOST_polar_frac}
    \end{figure}

This overall failure to find a good model reproducing the observations is not particularly surprising given that no Mie or DHS models were able to explain the unpolarised SPF.  This is why \citet{Milli2017} suggested the presence of aggregates to reproduce the large peak of forward scattering detected in unpolarised NIR light, and the slight backward scattering behaviour. 

\subsection{Alternative scenarios: irregular surface roughness or fluffy aggregates}

It is interesting to note that the polarimetric properties of dust aggregates tend to be similar to those of their individual monomers, while these aggregates behave like large particles in unpolarised light \citep{Kataoka2014,Min2016,Olofsson2016}. This could explain why the best Mie models favour small particles while the unpolarised SPF in the H band suggests large particles. Despite tremendous progress in numerical tools to compute the scattering properties of large aggregates \cite[see][for a review]{Kolokolova2004}, it is still difficult to simulate a polydisperse mixture of irregular aggregated particles. In particular, it is very computationally demanding to consider particle size parameters $x\gg1$ (i.e. size $\gg$ wavelength). This modelling is outside the scope of this paper, but we present here some  studies that may have potential for future characterisation of the scattering properties of the particles. \citet{Kolokolova2015} introduced a tool to simulate a polydisperse mixture of randomly oriented smooth and rough spheroids of a variety of aspect ratios. These latter authors were able to reproduce the main photopolarimetric characteristics of the light scattered by cometary dust. The pSPF of their best rough spheroid model does not however fit the pSPF measured on HR\,4796 with ZIMPOL, because their pSPF continuously increases from $160^\circ$ to $10^\circ$ scattering angle, unlike our data which plateaus between $30^\circ$ and $60^\circ$. To simulate aggregates, \citet{Min2016} used the discrete dipole approximation to efficiently compute the scattering properties of compact aggregates while \citet{Tazaki2016} showed that the Rayleigh-Gans-Debye theory can be efficiently used to approach the behaviour of fractal aggregates. 

\begin{figure}
   \centering
   \includegraphics[width=\hsize]{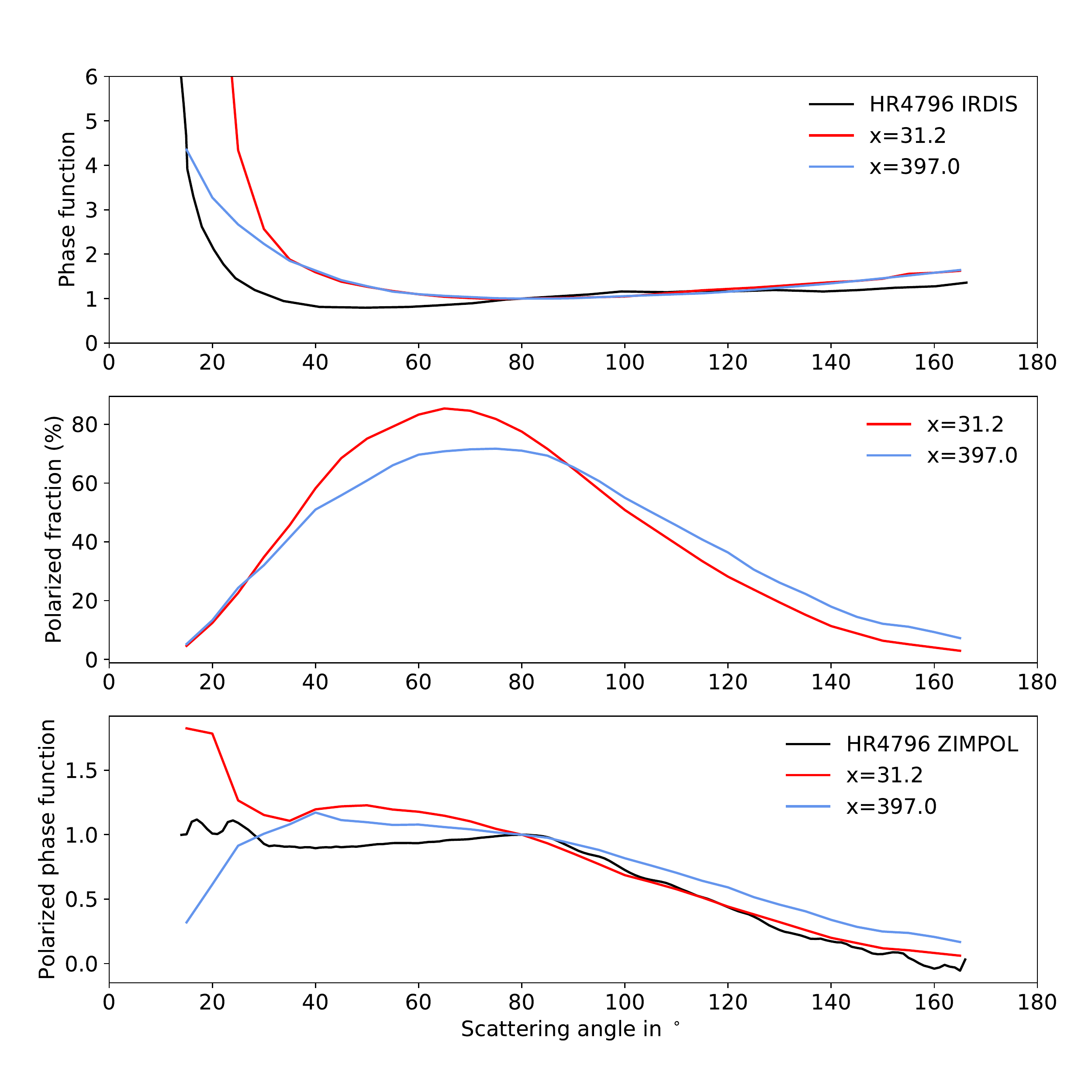}
   \caption{Unpolarised SPF (top), polarised fraction (middle), and polarised SPF (bottom) of two models of compact spheres with a rough irregular surface \citep{Mukai1982}, with size parameters $x$ of 31 (red curve) and 392 (blue curve). For the unpolarised SPF (top graph), we overplotted that derived for HR\,4796 in the NIR at 1.6\micron{} (black curve). At this wavelength, the red and blue model SPFs correspond to particle sizes of 8 and 101\micron, respectively. For the polarised SPF (bottom graph) we overplotted that of HR\,4796 as derived in the optical in this study (black curve). At this wavelength, the red and blue model SPFs correspond to particle sizes of 3.7 and 47\micron, respectively.}
    \label{fig_comparison_Mukai}
    \end{figure}

Another scenario to explain both the unpolarised and polarised SPF would be the presence of large particles with a random rough surface, as modelled in \citet{Mukai1982}. They show two models of large absorbing particles with surface roughness and size parameter $x=31.2$ and $x=397$ that convincingly match measurements from laboratory microwave analogues, and that we reproduced in Fig. \ref{fig_comparison_Mukai}.  These two size parameters correspond to particles of  3.7\micron{} and 46.5\micron{} in size,
respectively, at the wavelength of the ZIMPOL observations, and to particles of 7.9\micron{} and 101\micron{} in the H-band. As shown in Fig. \ref{fig_comparison_Mukai} (top curve), these models can reproduce the overall shape of the unpolarised SPF, with a peak of forward scattering and a mild backward scattering behaviour, although the location of the peak does not match the H-band measurements of \citet{Milli2017} because the size parameter is likely in between the two models.
This model is nonetheless instructive to understand in which case the polarised SPF can increase at small scattering angles. The polarised SPF is the product of the unpolarised SPF and polarised fraction, and these two functions show competing behaviours: the unpolarised SPF peaks at short scattering angles, while the polarised fraction goes down to zero. Therefore, in the range of scattering angles around $20^\circ$ which is of interest for this study, the polarised phase function can either decrease, as is the case for $x=397$, or increase,  as is the case for $x=31.2$. From our ZIMPOL measurement, we are probably in an intermediate case, as shown in Fig. \ref{fig_comparison_Mukai} (bottom curve), with an overall particle size of $\sim$20\micron{}. This model can also explain why the contrast between the forward-scattering peak and the ansae is higher in the Ks band than in the optical, as seen by comparing the ZIMPOL image with that of \citet{Perrin2015} in polarized light. The size parameter decreases when the wavelength increases, and the polarised SPF increases at small scattering angles (red curve in Fig. \ref{fig_comparison_Mukai} bottom). 

These two scenarios differ in the compactness of the particles. In the first case, large fluffy aggregates are needed to reproduce the unpolarised SPF, while in the second case the particles are compact and show surface roughness.  

\subsection{Comparison to solar-system comets}

Comets in our solar system can inform us of the dust properties of debris discs, as cold debris rings are considered as a reservoir of cometary material releasing smaller particles through collisions \cite[see][for a review]{Kral2018_prospective,Hughes2018}.

Unfortunately, no comets have yet been characterised at such small scattering angles because of the difficulty in disentangling direct sunlight from dust-scattered light. Recently, the SPF of the comet 67P could be retrieved from $25^\circ$ to $165^\circ$ \citep{Bertini2017} thanks to the Rosetta mission, and this represents to our knowledge the widest range available. The measurement of the unpolarised SPF shows a strong backward scattering behaviour, not compatible with the SPF of HR\,4796 as measured in the H-band. 
The polarised fraction is a common diagnostic tool to classify comets, and this has been documented so far for scattering angles between $28^\circ$ and $180^\circ$ \citep{Kolokolova2004}. It has been used to classify comets in three classes depending on their maximum polarisation fraction: low-polarisation comets (10-15\%) and higher-polarisation comets (25-30\%), with a separate class for a few notable exceptions with higher polarisation levels such as C/1995 O1 (Hale-Bopp). HR\,4796 would therefore belong to this latter class, with a maximum polarised fraction of $50\% \pm 3\%$ in the Ks band \citep{Perrin2015} and  40\% $\pm$ 26\% in the optical at $90^\circ$ scattering angle. Similar or higher levels of polarisation were also detected in the edge-on disc AU Mic \citep{Graham2007}, where very porous aggregates ($>70\%$ porosity) were proposed to reproduce both the polarised fraction and unpolarised SPF. 
For both comets and debris discs, the absence of resonant oscillations in the SPF is the main reason why the geometry of the particles is believed to differ significantly from that of a perfect sphere \citep{Kolokolova2004}. 
The high degree of polarisation for comet C/1995 O1 (Hale-Bopp) was interpreted as the presence of small particles below 1\micron{} included in fluffy aggregates \citep{Hadamcik2003,Levasseur-Regourd2007}. The fact that our Mie and DHS modelling of the polarised SPF point towards submicronic particles could indicate similar particles are present in the HR\,4796 ring, likely part of larger aggregates to avoid rapid blow-out by the radiation pressure. 

All comets show a negative branch of polarisation between scattering angles of $158^\circ$ and $180^\circ$, with a minimum of about -1.5\% \citep{Kiselev2015}. This signal would also appear as a negative signal in our $Q_\phi$ image but we do not have the sensitivity to detect it in this data set. The linear trend between $100$ and $140^\circ$ suggests however an inversion around $160^\circ$, compatible with cometary material.

A deeper dataset in the absence of LWE would reveal this critical region of the disc. In addition, a simultaneous measurement of the unpolarised intensity of the disc would make it possible to retrieve the polarised fraction of the dust along the ring, for a direct comparison to comets in the solar system. This is possible with the ZIMPOL instruments with either a dedicated pupil-stabilised data set in unpolarised light, or the so-called p1 mode where the derotator is fixed to minimise instrumental polarisation.

\section{Conclusions}
\label{sec_conclusions}

The polarised SPF of the dust around HR\,4796 was measured for the first time in the optical from $13^\circ$ up to $\sim145^\circ$ with the SPHERE/ZIMPOL instrument. The SPF starts by decreasing beyond $13^\circ$ to plateau between $30^\circ$ and $80^\circ$, and then decreases again linearly.  The sensitivity of the data is not high enough to measure an inversion in the polarised fraction for large scattering angles, but the trend suggests that the polarised fraction would cancel at about $160^\circ$, as is the case for all comets of our solar system.
The overall behaviour of the polarised scattering phase function is hardly compatible with compact Mie or DHS spheres. Those two theories suggest predominance of small sub-micronic particles which would be rapidly blown out of the system by the radiation pressure of the central star. In addition, no resonant oscillations typical for spherical particles are visible in the polarised scattering phase function. The particles are therefore more complex than perfect spheres, as already concluded from previous studies. Large fluffy porous particles could explain the polarised properties if the individual monomers are small enough; large compact particles with irregular surface roughness could be an alternative solution. However, further modelling work is required to validate those scenarios, as most available models, tailored for solar-system comets, focus on reproducing the maximum/minimum polarised fraction and their locations, which are not available in these observations. Extracting the unpolarised phase function and polarised fraction simultaneously in the optical is therefore a logical next step to further explore the analogy with comets. 



%
%
%
%

\begin{acknowledgements}
J.M. thanks the ESO Office for Science for making the use of the Chapman computer server possible, to run the MCMC analysis presented in Appendix \ref{App_convolution}. F.M. acknowledges funding from ANR of France under contract number ANR-16-CE31-0013. J.O. acknowledges support from the ICM (Iniciativa Cient\'ifica Milenio) via the Nucleo Milenio de Formaci\'on planetaria grant, from the Universidad de Valpara\'iso and from Fondecyt (grant 1180395). A.Z. acknowledges support from the CONICYT + PAI/ Convocatoria nacional subvenci\'on a la instalaci\'on en la academia, convocatoria 2017 + Folio PAI77170087. 
SPHERE is an instrument designed and built by a consortium
consisting of IPAG (Grenoble, France), MPIA (Heidelberg, Germany), LAM
(Marseille, France), LESIA (Paris, France), Laboratoire Lagrange
(Nice, France), INAF Osservatorio di Padova (Italy), Observatoire de
Gen\`eve (Switzerland), ETH Zurich (Switzerland), NOVA (Netherlands),
ONERA (France) and ASTRON (Netherlands) in collaboration with
 ESO. SPHERE was funded by ESO, with additional contributions from CNRS
(France), MPIA (Germany), INAF (Italy), FINES (Switzerland) and NOVA
 (Netherlands).  SPHERE also received funding from the European
 Commission Sixth and Seventh Framework Programmes as part of the
 Optical Infrared Coordination Network for Astronomy (OPTICON) under
 grant number RII3-Ct-2004-001566 for FP6 (2004?2008), grant number
 226604 for FP7 (2009?2012) and grant number 312430 for FP7
 (2013/2016). We also acknowledge financial support from the Programme National de
 Plan\'etologie (PNP) and the Programme National de Physique Stellaire
 (PNPS) of CNRS-INSU in France. This work has also been supported by a grant from
 the French Labex OSUG@2020 (Investissements d'avenir ANR10 LABX56).
 The project is supported by CNRS, by the Agence Nationale de la
 Recherche (ANR-14-CE33-0018). It has also been carried out within the framework of the National Centre for Competence in 
 Research PlanetS supported by the Swiss National Science Foundation (SNSF). HMS is pleased 
 to acknowledge this financial support of the SNSF. This research made use of Astropy\footnote{http://www.astropy.org} a community-developed core Python package for Astronomy \citep{Astropy2013,Astropy2018}.

\end{acknowledgements}

\bibliography{biblio_HR4796_zimpol}     

\newcommand{\noop}[1]{}
\begin{thebibliography}{57}
\expandafter\ifx\csname natexlab\endcsname\relax\def\natexlab#1{#1}\fi

\bibitem[{{Arnold} {et~al.}(2019){Arnold}, {Weinberger}, {Videen}, \&
  {Zubko}}]{Arnold2019}
{Arnold}, J.~A., {Weinberger}, A.~J., {Videen}, G., \& {Zubko}, E.~S. 2019,
  arXiv e-prints [\eprint[arXiv]{1902.10183}]

\bibitem[{{Astropy Collaboration} {et~al.}(2018){Astropy Collaboration},
  {Price-Whelan}, {Sip{\H o}cz}, {G{\"u}nther}, {Lim}, {Crawford}, {Conseil},
  {Shupe}, {Craig}, {Dencheva}, {Ginsburg}, {VanderPlas}, {Bradley},
  {P{\'e}rez-Su{\'a}rez}, {de Val-Borro}, {Aldcroft}, {Cruz}, {Robitaille},
  {Tollerud}, {Ardelean}, {Babej}, {Bach}, {Bachetti}, {Bakanov}, {Bamford},
  {Barentsen}, {Barmby}, {Baumbach}, {Berry}, {Biscani}, {Boquien}, {Bostroem},
  {Bouma}, {Brammer}, {Bray}, {Breytenbach}, {Buddelmeijer}, {Burke},
  {Calderone}, {Cano Rodr{\'{\i}}guez}, {Cara}, {Cardoso}, {Cheedella},
  {Copin}, {Corrales}, {Crichton}, {D'Avella}, {Deil}, {Depagne}, {Dietrich},
  {Donath}, {Droettboom}, {Earl}, {Erben}, {Fabbro}, {Ferreira}, {Finethy},
  {Fox}, {Garrison}, {Gibbons}, {Goldstein}, {Gommers}, {Greco}, {Greenfield},
  {Groener}, {Grollier}, {Hagen}, {Hirst}, {Homeier}, {Horton}, {Hosseinzadeh},
  {Hu}, {Hunkeler}, {Ivezi{\'c}}, {Jain}, {Jenness}, {Kanarek}, {Kendrew},
  {Kern}, {Kerzendorf}, {Khvalko}, {King}, {Kirkby}, {Kulkarni}, {Kumar},
  {Lee}, {Lenz}, {Littlefair}, {Ma}, {Macleod}, {Mastropietro}, {McCully},
  {Montagnac}, {Morris}, {Mueller}, {Mumford}, {Muna}, {Murphy}, {Nelson},
  {Nguyen}, {Ninan}, {N{\"o}the}, {Ogaz}, {Oh}, {Parejko}, {Parley}, {Pascual},
  {Patil}, {Patil}, {Plunkett}, {Prochaska}, {Rastogi}, {Reddy Janga},
  {Sabater}, {Sakurikar}, {Seifert}, {Sherbert}, {Sherwood-Taylor}, {Shih},
  {Sick}, {Silbiger}, {Singanamalla}, {Singer}, {Sladen}, {Sooley},
  {Sornarajah}, {Streicher}, {Teuben}, {Thomas}, {Tremblay}, {Turner},
  {Terr{\'o}n}, {van Kerkwijk}, {de la Vega}, {Watkins}, {Weaver}, {Whitmore},
  {Woillez}, {Zabalza}, \& {Astropy Contributors}}]{Astropy2018}
{Astropy Collaboration}, {Price-Whelan}, A.~M., {Sip{\H o}cz}, B.~M., {et~al.}
  2018, \aj, 156, 123

\bibitem[{{Astropy Collaboration} {et~al.}(2013){Astropy Collaboration},
  {Robitaille}, {Tollerud}, {Greenfield}, {Droettboom}, {Bray}, {Aldcroft},
  {Davis}, {Ginsburg}, {Price-Whelan}, {Kerzendorf}, {Conley}, {Crighton},
  {Barbary}, {Muna}, {Ferguson}, {Grollier}, {Parikh}, {Nair}, {Unther},
  {Deil}, {Woillez}, {Conseil}, {Kramer}, {Turner}, {Singer}, {Fox}, {Weaver},
  {Zabalza}, {Edwards}, {Azalee Bostroem}, {Burke}, {Casey}, {Crawford},
  {Dencheva}, {Ely}, {Jenness}, {Labrie}, {Lim}, {Pierfederici}, {Pontzen},
  {Ptak}, {Refsdal}, {Servillat}, \& {Streicher}}]{Astropy2013}
{Astropy Collaboration}, {Robitaille}, T.~P., {Tollerud}, E.~J., {et~al.} 2013,
  \aap, 558, A33

\bibitem[{{Augereau} {et~al.}(1999){Augereau}, {Lagrange}, {Mouillet},
  {Papaloizou}, \& {Grorod}}]{Augereau1999}
{Augereau}, J.~C., {Lagrange}, A.~M., {Mouillet}, D., {Papaloizou}, J.~C.~B.,
  \& {Grorod}, P.~A. 1999, \aap, 348, 557

\bibitem[{{Bell} {et~al.}(2015){Bell}, {Mamajek}, \& {Naylor}}]{Bell2015}
{Bell}, C.~P.~M., {Mamajek}, E.~E., \& {Naylor}, T. 2015, \mnras, 454, 593

\bibitem[{{Bertini} {et~al.}(2017){Bertini}, {La Forgia}, {Tubiana},
  {G{\"u}ttler}, {Fulle}, {Moreno}, {Frattin}, {Kovacs}, {Pajola}, {Sierks},
  {Barbieri}, {Lamy}, {Rodrigo}, {Koschny}, {Rickman}, {Keller}, {Agarwal},
  {A'Hearn}, {Barucci}, {Bertaux}, {Bodewits}, {Cremonese}, {Da Deppo},
  {Davidsson}, {Debei}, {De Cecco}, {Drolshagen}, {Ferrari}, {Ferri},
  {Fornasier}, {Gicquel}, {Groussin}, {Gutierrez}, {Hasselmann}, {Hviid}, {Ip},
  {Jorda}, {Knollenberg}, {Kramm}, {K{\"u}hrt}, {K{\"u}ppers}, {Lara},
  {Lazzarin}, {Lin}, {Moreno}, {Lucchetti}, {Marzari}, {Massironi}, {Mottola},
  {Naletto}, {Oklay}, {Ott}, {Penasa}, {Thomas}, \& {Vincent}}]{Bertini2017}
{Bertini}, I., {La Forgia}, F., {Tubiana}, C., {et~al.} 2017, \mnras, 469, S404

\bibitem[{{Beuzit} {et~al.}(2019){Beuzit}, {Vigan}, {Mouillet}, {Dohlen},
  {Gratton}, {Boccaletti}, {Sauvage}, {Schmid}, {Langlois}, {Petit},
  {Baruffolo}, {Feldt}, {Milli}, {Wahhaj}, {Abe}, {Anselmi}, {Antichi},
  {Barette}, {Baudrand}, {Baudoz}, {Bazzon}, {Bernardi}, {Blanchard}, {Brast},
  {Bruno}, {Buey}, {Carbillet}, {Carle}, {Cascone}, {Chapron}, {Chauvin},
  {Charton}, {Claudi}, {Costille}, {De Caprio}, {Delboulb{\'e}}, {Desidera},
  {Dominik}, {Downing}, {Dupuis}, {Fabron}, {Fantinel}, {Farisato},
  {Feautrier}, {Fedrigo}, {Fusco}, {Gigan}, {Ginski}, {Girard}, {Giro},
  {Gisler}, {Gluck}, {Gry}, {Henning}, {Hubin}, {Hugot}, {Incorvaia}, {Jaquet},
  {Kasper}, {Lagadec}, {Lagrange}, {Le Coroller}, {Le Mignant}, {Le Ruyet},
  {Lessio}, {Lizon}, {Llored}, {Lundin}, {Madec}, {Magnard}, {Marteaud},
  {Martinez}, {Maurel}, {M{\'e}nard}, {Mesa}, {M{\"o}ller-Nilsson}, {Moulin},
  {Moutou}, {Orign{\'e}}, {Parisot}, {Pavlov}, {Perret}, {Pragt}, {Puget},
  {Rabou}, {Ramos}, {Reess}, {Rigal}, {Rochat}, {Roelfsema}, {Rousset}, {Roux},
  {Saisse}, {Salasnich}, {Santambrogio}, {Scuderi}, {Segransan}, {Sevin},
  {Siebenmorgen}, {Soenke}, {Stadler}, {Suarez}, {Tiph{\`e}ne}, {Turatto},
  {Udry}, {Vakili}, {Waters}, {Weber}, {Wildi}, {Zins}, \&
  {Zurlo}}]{Beuzit2019}
{Beuzit}, J.-L., {Vigan}, A., {Mouillet}, D., {et~al.} 2019, arXiv e-prints
  [\eprint[arXiv]{1902.04080}]

\bibitem[{{Engler} {et~al.}(2018){Engler}, {Schmid}, {Quanz}, {Avenhaus}, \&
  {Bazzon}}]{Engler2018}
{Engler}, N., {Schmid}, H.~M., {Quanz}, S.~P., {Avenhaus}, H., \& {Bazzon}, A.
  2018, \aap, 618, A151

\bibitem[{{Engler} {et~al.}(2017){Engler}, {Schmid}, {Thalmann}, {Boccaletti},
  {Bazzon}, {Baruffolo}, {Beuzit}, {Claudi}, {Costille}, {Desidera}, {Dohlen},
  {Dominik}, {Feldt}, {Fusco}, {Ginski}, {Gisler}, {Girard}, {Gratton},
  {Henning}, {Hubin}, {Janson}, {Kasper}, {Kral}, {Langlois}, {Lagadec},
  {M{\'e}nard}, {Meyer}, {Milli}, {Mouillet}, {Olofsson}, {Pavlov}, {Pragt},
  {Puget}, {Quanz}, {Roelfsema}, {Salasnich}, {Siebenmorgen}, {Sissa},
  {Suarez}, {Szulagyi}, {Turatto}, {Udry}, \& {Wildi}}]{Engler2017}
{Engler}, N., {Schmid}, H.~M., {Thalmann}, C., {et~al.} 2017, \aap, 607, A90

\bibitem[{{Foreman-Mackey} {et~al.}(2013){Foreman-Mackey}, {Hogg}, {Lang}, \&
  {Goodman}}]{Foreman-Mackey2013}
{Foreman-Mackey}, D., {Hogg}, D.~W., {Lang}, D., \& {Goodman}, J. 2013, \pasp,
  125, 306

\bibitem[{{Gaia Collaboration} {et~al.}(2018){Gaia Collaboration}, {Brown},
  {Vallenari}, {Prusti}, {de Bruijne}, {Babusiaux}, {Bailer-Jones}, {Biermann},
  {Evans}, {Eyer}, {Jansen}, {Jordi}, {Klioner}, {Lammers}, {Lindegren},
  {Luri}, {Mignard}, {Panem}, {Pourbaix}, {Randich}, {Sartoretti}, {Siddiqui},
  {Soubiran}, {van Leeuwen}, {Walton}, {Arenou}, {Bastian}, {Cropper},
  {Drimmel}, {Katz}, {Lattanzi}, {Bakker}, {Cacciari}, {Casta{\~n}eda},
  {Chaoul}, {Cheek}, {De Angeli}, {Fabricius}, {Guerra}, {Holl}, {Masana},
  {Messineo}, {Mowlavi}, {Nienartowicz}, {Panuzzo}, {Portell}, {Riello},
  {Seabroke}, {Tanga}, {Th{\'e}venin}, {Gracia-Abril}, {Comoretto},
  {Garcia-Reinaldos}, {Teyssier}, {Altmann}, {Andrae}, {Audard},
  {Bellas-Velidis}, {Benson}, {Berthier}, {Blomme}, {Burgess}, {Busso},
  {Carry}, {Cellino}, {Clementini}, {Clotet}, {Creevey}, {Davidson}, {De
  Ridder}, {Delchambre}, {Dell'Oro}, {Ducourant}, {Fern{\'a}ndez-
  Hern{\'a}ndez}, {Fouesneau}, {Fr{\'e}mat}, {Galluccio}, {Garc{\'\i}a-Torres},
  {Gonz{\'a}lez-N{\'u}{\~n}ez}, {Gonz{\'a}lez-Vidal}, {Gosset}, {Guy},
  {Halbwachs}, {Hambly}, {Harrison}, {Hern{\'a}ndez}, {Hestroffer}, {Hodgkin},
  {Hutton}, {Jasniewicz}, {Jean-Antoine-Piccolo}, {Jordan}, {Korn},
  {Krone-Martins}, {Lanzafame}, {Lebzelter}, {L{\"o}ffler}, {Manteiga},
  {Marrese}, {Mart{\'\i}n-Fleitas}, {Moitinho}, {Mora}, {Muinonen}, {Osinde},
  {Pancino}, {Pauwels}, {Petit}, {Recio-Blanco}, {Richards}, {Rimoldini},
  {Robin}, {Sarro}, {Siopis}, {Smith}, {Sozzetti}, {S{\"u}veges}, {Torra}, {van
  Reeven}, {Abbas}, {Abreu Aramburu}, {Accart}, {Aerts}, {Altavilla},
  {{\'A}lvarez}, {Alvarez}, {Alves}, {Anderson}, {Andrei}, {Anglada Varela},
  {Antiche}, {Antoja}, {Arcay}, {Astraatmadja}, {Bach}, {Baker},
  {Balaguer-N{\'u}{\~n}ez}, {Balm}, {Barache}, {Barata}, {Barbato}, {Barblan},
  {Barklem}, {Barrado}, {Barros}, {Barstow}, {Bartholom{\'e} Mu{\~n}oz},
  {Bassilana}, {Becciani}, {Bellazzini}, {Berihuete}, {Bertone}, {Bianchi},
  {Bienaym{\'e}}, {Blanco-Cuaresma}, {Boch}, {Boeche}, {Bombrun}, {Borrachero},
  {Bossini}, {Bouquillon}, {Bourda}, {Bragaglia}, {Bramante}, {Breddels},
  {Bressan}, {Brouillet}, {Br{\"u}semeister}, {Brugaletta}, {Bucciarelli},
  {Burlacu}, {Busonero}, {Butkevich}, {Buzzi}, {Caffau}, {Cancelliere},
  {Cannizzaro}, {Cantat-Gaudin}, {Carballo}, {Carlucci}, {Carrasco},
  {Casamiquela}, {Castellani}, {Castro-Ginard}, {Charlot}, {Chemin},
  {Chiavassa}, {Cocozza}, {Costigan}, {Cowell}, {Crifo}, {Crosta}, {Crowley},
  {Cuypers}, {Dafonte}, {Damerdji}, {Dapergolas}, {David}, {David}, {de
  Laverny}, {De Luise}, {De March}, {de Martino}, {de Souza}, {de Torres},
  {Debosscher}, {del Pozo}, {Delbo}, {Delgado}, {Delgado}, {Di Matteo},
  {Diakite}, {Diener}, {Distefano}, {Dolding}, {Drazinos}, {Dur{\'a}n},
  {Edvardsson}, {Enke}, {Eriksson}, {Esquej}, {Eynard Bontemps}, {Fabre},
  {Fabrizio}, {Faigler}, {Falc{\~a}o}, {Farr{\`a}s Casas}, {Federici},
  {Fedorets}, {Fernique}, {Figueras}, {Filippi}, {Findeisen}, {Fonti},
  {Fraile}, {Fraser}, {Fr{\'e}zouls}, {Gai}, {Galleti}, {Garabato},
  {Garc{\'\i}a-Sedano}, {Garofalo}, {Garralda}, {Gavel}, {Gavras}, {Gerssen},
  {Geyer}, {Giacobbe}, {Gilmore}, {Girona}, {Giuffrida}, {Glass}, {Gomes},
  {Granvik}, {Gueguen}, {Guerrier}, {Guiraud}, {Guti{\'e}rrez-S{\'a}nchez},
  {Haigron}, {Hatzidimitriou}, {Hauser}, {Haywood}, {Heiter}, {Helmi}, {Heu},
  {Hilger}, {Hobbs}, {Hofmann}, {Holland}, {Huckle}, {Hypki}, {Icardi},
  {Jan{\ss}en}, {Jevardat de Fombelle}, {Jonker}, {Juh{\'a}sz}, {Julbe},
  {Karampelas}, {Kewley}, {Klar}, {Kochoska}, {Kohley}, {Kolenberg},
  {Kontizas}, {Kontizas}, {Koposov}, {Kordopatis}, {Kostrzewa-Rutkowska},
  {Koubsky}, {Lambert}, {Lanza}, {Lasne}, {Lavigne}, {Le Fustec}, {Le
  Poncin-Lafitte}, {Lebreton}, {Leccia}, {Leclerc}, {Lecoeur-Taibi},
  {Lenhardt}, {Leroux}, {Liao}, {Licata}, {Lindstr{\o}m}, {Lister}, {Livanou},
  {Lobel}, {L{\'o}pez}, {Managau}, {Mann}, {Mantelet}, {Marchal}, {Marchant},
  {Marconi}, {Marinoni}, {Marschalk{\'o}}, {Marshall}, {Martino}, {Marton},
  {Mary}, {Massari}, {Matijevi{\v{c}}}, {Mazeh}, {McMillan}, {Messina},
  {Michalik}, {Millar}, {Molina}, {Molinaro}, {Moln{\'a}r}, {Montegriffo},
  {Mor}, {Morbidelli}, {Morel}, {Morris}, {Mulone}, {Muraveva}, {Musella},
  {Nelemans}, {Nicastro}, {Noval}, {O'Mullane}, {Ord{\'e}novic},
  {Ord{\'o}{\~n}ez-Blanco}, {Osborne}, {Pagani}, {Pagano}, {Pailler},
  {Palacin}, {Palaversa}, {Panahi}, {Pawlak}, {Piersimoni}, {Pineau}, {Plachy},
  {Plum}, {Poggio}, {Poujoulet}, {Pr{\v{s}}a}, {Pulone}, {Racero}, {Ragaini},
  {Rambaux}, {Ramos-Lerate}, {Regibo}, {Reyl{\'e}}, {Riclet}, {Ripepi}, {Riva},
  {Rivard}, {Rixon}, {Roegiers}, {Roelens}, {Romero-G{\'o}mez}, {Rowell},
  {Royer}, {Ruiz-Dern}, {Sadowski}, {Sagrist{\`a} Sell{\'e}s}, {Sahlmann},
  {Salgado}, {Salguero}, {Sanna}, {Santana- Ros}, {Sarasso}, {Savietto},
  {Schultheis}, {Sciacca}, {Segol}, {Segovia}, {S{\'e}gransan}, {Shih},
  {Siltala}, {Silva}, {Smart}, {Smith}, {Solano}, {Solitro}, {Sordo}, {Soria
  Nieto}, {Souchay}, {Spagna}, {Spoto}, {Stampa}, {Steele},
  {Steidelm{\"u}ller}, {Stephenson}, {Stoev}, {Suess}, {Surdej}, {Szabados},
  {Szegedi-Elek}, {Tapiador}, {Taris}, {Tauran}, {Taylor}, {Teixeira},
  {Terrett}, {Teyssandier}, {Thuillot}, {Titarenko}, {Torra Clotet}, {Turon},
  {Ulla}, {Utrilla}, {Uzzi}, {Vaillant}, {Valentini}, {Valette}, {van Elteren},
  {Van Hemelryck}, {van Leeuwen}, {Vaschetto}, {Vecchiato}, {Veljanoski},
  {Viala}, {Vicente}, {Vogt}, {von Essen}, {Voss}, {Votruba}, {Voutsinas},
  {Walmsley}, {Weiler}, {Wertz}, {Wevers}, {Wyrzykowski}, {Yoldas},
  {{\v{Z}}erjal}, {Ziaeepour}, {Zorec}, {Zschocke}, {Zucker}, {Zurbach}, \&
  {Zwitter}}]{Gaia2018}
{Gaia Collaboration}, {Brown}, A.~G.~A., {Vallenari}, A., {et~al.} 2018, \aap,
  616, A1

\bibitem[{{Gomez Gonzalez} {et~al.}(2017){Gomez Gonzalez}, {Wertz}, {Absil},
  {Christiaens}, {Defr{\`e}re}, {Mawet}, {Milli}, {Absil}, {Van Droogenbroeck},
  {Cantalloube}, {Hinz}, {Skemer}, {Karlsson}, \& {Surdej}}]{Gomez2017}
{Gomez Gonzalez}, C.~A., {Wertz}, O., {Absil}, O., {et~al.} 2017, \aj, 154, 7

\bibitem[{{Graham} {et~al.}(2007){Graham}, {Kalas}, \& {Matthews}}]{Graham2007}
{Graham}, J.~R., {Kalas}, P.~G., \& {Matthews}, B.~C. 2007, \apj, 654, 595

\bibitem[{{Hadamcik} \& {Levasseur-Regourd}(2003)}]{Hadamcik2003}
{Hadamcik}, E. \& {Levasseur-Regourd}, A.~C. 2003, \aap, 403, 757

\bibitem[{{Henyey} \& {Greenstein}(1941)}]{Henyey1941}
{Henyey}, L.~G. \& {Greenstein}, J.~L. 1941, \apj, 93, 70

\bibitem[{{H{\o}g} {et~al.}(2000){H{\o}g}, {Fabricius}, {Makarov}, {Urban},
  {Corbin}, {Wycoff}, {Bastian}, {Schwekendiek}, \& {Wicenec}}]{Hog2000}
{H{\o}g}, E., {Fabricius}, C., {Makarov}, V.~V., {et~al.} 2000, \aap, 355, L27

\bibitem[{{Hughes} {et~al.}(2018){Hughes}, {Duch{\^e}ne}, \&
  {Matthews}}]{Hughes2018}
{Hughes}, A.~M., {Duch{\^e}ne}, G., \& {Matthews}, B.~C. 2018, Annual Review of
  Astronomy and Astrophysics, 56, 541

\bibitem[{{Kataoka} {et~al.}(2014){Kataoka}, {Okuzumi}, {Tanaka}, \&
  {Nomura}}]{Kataoka2014}
{Kataoka}, A., {Okuzumi}, S., {Tanaka}, H., \& {Nomura}, H. 2014, \aap, 568,
  A42

\bibitem[{{Kennedy} {et~al.}(2018){Kennedy}, {Marino}, {Matr{\`a}},
  {Pani{\'c}}, {Wilner}, {Wyatt}, \& {Yelverton}}]{Kennedy2018_HR4796}
{Kennedy}, G.~M., {Marino}, S., {Matr{\`a}}, L., {et~al.} 2018, \mnras, 475,
  4924

\bibitem[{{Kiselev} {et~al.}(2015){Kiselev}, {Rosenbush}, {Levasseur-Regourd},
  \& {Kolokolova}}]{Kiselev2015}
{Kiselev}, N., {Rosenbush}, V., {Levasseur-Regourd}, A.-C., \& {Kolokolova}, L.
  2015, {Comets}, 379

\bibitem[{{Kolokolova} {et~al.}(2015){Kolokolova}, {Das}, {Dubovik},
  {Lapyonok}, \& {Yang}}]{Kolokolova2015}
{Kolokolova}, L., {Das}, H.~S., {Dubovik}, O., {Lapyonok}, T., \& {Yang}, P.
  2015, Planetary and Space Science, 116, 30

\bibitem[{{Kolokolova} {et~al.}(2004){Kolokolova}, {Hanner},
  {Levasseur-Regourd}, \& {Gustafson}}]{Kolokolova2004}
{Kolokolova}, L., {Hanner}, M.~S., {Levasseur-Regourd}, A.~C., \& {Gustafson},
  B. {\r{A}}.~S. 2004, {Physical properties of cometary dust from light
  scattering and thermal emission}, 577

\bibitem[{{Kral} {et~al.}(2018){Kral}, {Clarke}, \&
  {Wyatt}}]{Kral2018_prospective}
{Kral}, Q., {Clarke}, C., \& {Wyatt}, M.~C. 2018, {Circumstellar Discs: What
  Will Be Next?}, 165

\bibitem[{{Lagrange} {et~al.}(2012){Lagrange}, {Milli}, {Boccaletti}, {Lacour},
  {Thebault}, {Chauvin}, {Mouillet}, {Augereau}, {Bonnefoy}, {Ehrenreich}, \&
  {Kral}}]{Lagrange2012_HR4796}
{Lagrange}, A.-M., {Milli}, J., {Boccaletti}, A., {et~al.} 2012, \aap, 546, A38

\bibitem[{{Levasseur-Regourd} {et~al.}(2007){Levasseur-Regourd}, {Mukai},
  {Lasue}, \& {Okada}}]{Levasseur-Regourd2007}
{Levasseur-Regourd}, A.~C., {Mukai}, T., {Lasue}, J., \& {Okada}, Y. 2007,
  Planetary and Space Science, 55, 1010

\bibitem[{{Marois} {et~al.}(2006){Marois}, {Lafreni{\`e}re}, {Doyon},
  {Macintosh}, \& {Nadeau}}]{Marois2006}
{Marois}, C., {Lafreni{\`e}re}, D., {Doyon}, R., {Macintosh}, B., \& {Nadeau},
  D. 2006, \apj, 641, 556

\bibitem[{{Matthews} {et~al.}(2014){Matthews}, {Krivov}, {Wyatt}, {Bryden}, \&
  {Eiroa}}]{Matthews2014}
{Matthews}, B.~C., {Krivov}, A.~V., {Wyatt}, M.~C., {Bryden}, G., \& {Eiroa},
  C. 2014, Protostars and Planets VI, 521

\bibitem[{{Milli} {et~al.}(2018){Milli}, {Kasper}, {Bourget}, {Pannetier},
  {Mouillet}, {Sauvage}, {Reyes}, {Fusco}, {Cantalloube}, {Tristam}, {Wahhaj},
  {Beuzit}, {Girard}, {Mawet}, {Telle}, {Vigan}, \& {N'Diaye}}]{Milli2018}
{Milli}, J., {Kasper}, M., {Bourget}, P., {et~al.} 2018, in Society of
  Photo-Optical Instrumentation Engineers (SPIE) Conference Series, Vol. 10703,
  107032A

\bibitem[{{Milli} {et~al.}(2015){Milli}, {Mawet}, {Pinte}, {Lagrange},
  {Mouillet}, {Girard}, {Augereau}, {De Boer}, {Pueyo}, \&
  {Choquet}}]{Milli2015}
{Milli}, J., {Mawet}, D., {Pinte}, C., {et~al.} 2015, \aap, 577, A57

\bibitem[{{Milli} {et~al.}(2012){Milli}, {Mouillet}, {Lagrange}, {Boccaletti},
  {Mawet}, {Chauvin}, \& {Bonnefoy}}]{Milli2012}
{Milli}, J., {Mouillet}, D., {Lagrange}, A.-M., {et~al.} 2012, \aap, 545, A111

\bibitem[{{Milli} {et~al.}(2017){Milli}, {Vigan}, {Mouillet}, {Lagrange},
  {Augereau}, {Pinte}, {Mawet}, {Schmid}, {Boccaletti}, {Matr{\`a}}, {Kral},
  {Ertel}, {Chauvin}, {Bazzon}, {M{\'e}nard}, {Beuzit}, {Thalmann}, {Dominik},
  {Feldt}, {Henning}, {Min}, {Girard}, {Galicher}, {Bonnefoy}, {Fusco}, {de
  Boer}, {Janson}, {Maire}, {Mesa}, {Schlieder}, \& {SPHERE
  Consortium}}]{Milli2017}
{Milli}, J., {Vigan}, A., {Mouillet}, D., {et~al.} 2017, \aap, 599, A108

\bibitem[{{Min} {et~al.}(2005){Min}, {Hovenier}, \& {de Koter}}]{Min2005}
{Min}, M., {Hovenier}, J.~W., \& {de Koter}, A. 2005, \aap, 432, 909

\bibitem[{{Min} {et~al.}(2016){Min}, {Rab}, {Woitke}, {Dominik}, \&
  {M{\'e}nard}}]{Min2016}
{Min}, M., {Rab}, C., {Woitke}, P., {Dominik}, C., \& {M{\'e}nard}, F. 2016,
  \aap, 585, A13

\bibitem[{{Moerchen} {et~al.}(2011){Moerchen}, {Churcher}, {Telesco}, {Wyatt},
  {Fisher}, \& {Packham}}]{Moerchen2011}
{Moerchen}, M.~M., {Churcher}, L.~J., {Telesco}, C.~M., {et~al.} 2011, \aap,
  526, A34

\bibitem[{{Mo{\'o}r} {et~al.}(2006){Mo{\'o}r}, {{\'A}brah{\'a}m}, {Derekas},
  {Kiss}, {Kiss}, {Apai}, {Grady}, \& {Henning}}]{Moor2006}
{Mo{\'o}r}, A., {{\'A}brah{\'a}m}, P., {Derekas}, A., {et~al.} 2006, \apj, 644,
  525

\bibitem[{{Mukai} {et~al.}(1982){Mukai}, {Mukai}, {Giese}, {Weiss}, \&
  {Zerull}}]{Mukai1982}
{Mukai}, S., {Mukai}, T., {Giese}, R.~H., {Weiss}, K., \& {Zerull}, R.~H. 1982,
  Moon and Planets, 26, 197

\bibitem[{{Olofsson} {et~al.}(2016){Olofsson}, {Samland}, {Avenhaus},
  {Caceres}, {Henning}, {Mo{\'o}r}, {Milli}, {Canovas}, {Quanz}, {Schreiber},
  {Augereau}, {Bayo}, {Bazzon}, {Beuzit}, {Boccaletti}, {Buenzli}, {Casassus},
  {Chauvin}, {Dominik}, {Desidera}, {Feldt}, {Gratton}, {Janson}, {Lagrange},
  {Langlois}, {Lannier}, {Maire}, {Mesa}, {Pinte}, {Rouan}, {Salter},
  {Thalmann}, \& {Vigan}}]{Olofsson2016}
{Olofsson}, J., {Samland}, M., {Avenhaus}, H., {et~al.} 2016, \aap, 591, A108

\bibitem[{{Pan} {et~al.}(2016){Pan}, {Nesvold}, \& {Kuchner}}]{Pan2016}
{Pan}, M., {Nesvold}, E.~R., \& {Kuchner}, M.~J. 2016, \apj, 832, 81

\bibitem[{{Perrin} {et~al.}(2015){Perrin}, {Duchene}, {Millar-Blanchaer},
  {Fitzgerald}, {Graham}, {Wiktorowicz}, {Kalas}, {Macintosh}, {Bauman},
  {Cardwell}, {Chilcote}, {De Rosa}, {Dillon}, {Doyon}, {Dunn}, {Erikson},
  {Gavel}, {Goodsell}, {Hartung}, {Hibon}, {Ingraham}, {Kerley}, {Konapacky},
  {Larkin}, {Maire}, {Marchis}, {Marois}, {Mittal}, {Morzinski}, {Oppenheimer},
  {Palmer}, {Patience}, {Poyneer}, {Pueyo}, {Rantakyr{\"o}}, {Sadakuni},
  {Saddlemyer}, {Savransky}, {Soummer}, {Sivaramakrishnan}, {Song}, {Thomas},
  {Wallace}, {Wang}, \& {Wolff}}]{Perrin2015}
{Perrin}, M.~D., {Duchene}, G., {Millar-Blanchaer}, M., {et~al.} 2015, \apj,
  799, 182

\bibitem[{{Pinte} {et~al.}(2006){Pinte}, {M{\'e}nard}, {Duch{\^e}ne}, \&
  {Bastien}}]{Pinte2006}
{Pinte}, C., {M{\'e}nard}, F., {Duch{\^e}ne}, G., \& {Bastien}, P. 2006, \aap,
  459, 797

\bibitem[{{Ray} \& {Srivastava}(2008)}]{Ray2008}
{Ray}, A. \& {Srivastava}, D.~C. 2008, Journal of Structural Geology, 30, 1593

\bibitem[{{Rodigas} {et~al.}(2014){Rodigas}, {Malhotra}, \&
  {Hinz}}]{Rodigas2014}
{Rodigas}, T.~J., {Malhotra}, R., \& {Hinz}, P.~M. 2014, \apj, 780, 65

\bibitem[{{Rodigas} {et~al.}(2015){Rodigas}, {Stark}, {Weinberger}, {Debes},
  {Hinz}, {Close}, {Chen}, {Smith}, {Males}, {Skemer}, {Puglisi}, {Follette},
  {Morzinski}, {Wu}, {Briguglio}, {Esposito}, {Pinna}, {Riccardi}, {Schneider},
  \& {Xompero}}]{Rodigas2015}
{Rodigas}, T.~J., {Stark}, C.~C., {Weinberger}, A., {et~al.} 2015, \apj, 798,
  96

\bibitem[{{Sauvage} {et~al.}(2016{\natexlab{a}}){Sauvage}, {Fusco}, {Lamb},
  {Girard}, {Brinkmann}, {Guesalaga}, {Wizinowich}, {O'Neal}, {N'Diaye},
  {Vigan}, {Mouillet}, {Beuzit}, {Kasper}, {Le Louarn}, {Milli}, {Dohlen},
  {Neichel}, {Bourget}, {Haguenauer}, \& {Mawet}}]{Sauvage2016}
{Sauvage}, J.-F., {Fusco}, T., {Lamb}, M., {et~al.} 2016{\natexlab{a}}, in
  \procspie, Vol. 9909, Adaptive Optics Systems V, 990916

\bibitem[{{Sauvage} {et~al.}(2016{\natexlab{b}}){Sauvage}, {Fusco}, {Petit},
  {Costille}, {Mouillet}, {Beuzit}, {Dohlen}, {Kasper}, {Suarez}, {Soenke},
  {Baruffolo}, {Salasnich}, {Rochat}, {Fedrigo}, {Baudoz}, {Hugot}, {Sevin},
  {Perret}, {Wildi}, {Downing}, {Feautrier}, {Puget}, {Vigan}, {O'Neal},
  {Girard}, {Mawet}, {Schmid}, \& {Roelfsema}}]{Sauvage2016_SAXO}
{Sauvage}, J.-F., {Fusco}, T., {Petit}, C., {et~al.} 2016{\natexlab{b}},
  Journal of Astronomical Telescopes, Instruments, and Systems, 2, 025003

\bibitem[{{Schmid} {et~al.}(2018){Schmid}, {Bazzon}, {Roelfsema}, {Mouillet},
  {Milli}, {Menard}, {Gisler}, {Hunziker}, {Pragt}, {Dominik}, {Boccaletti},
  {Ginski}, {Abe}, {Antoniucci}, {Avenhaus}, {Baruffolo}, {Baudoz}, {Beuzit},
  {Carbillet}, {Chauvin}, {Claudi}, {Costille}, {Daban}, {de Haan}, {Desidera},
  {Dohlen}, {Downing}, {Elswijk}, {Engler}, {Feldt}, {Fusco}, {Girard},
  {Gratton}, {Hanenburg}, {Henning}, {Hubin}, {Joos}, {Kasper}, {Keller},
  {Langlois}, {Lagadec}, {Martinez}, {Mulder}, {Pavlov}, {Podio}, {Puget},
  {Quanz}, {Rigal}, {Salasnich}, {Sauvage}, {Schuil}, {Siebenmorgen}, {Sissa},
  {Snik}, {Suarez}, {Thalmann}, {Turatto}, {Udry}, {van Duin}, {van Holstein},
  {Vigan}, \& {Wildi}}]{Schmid2018}
{Schmid}, H.~M., {Bazzon}, A., {Roelfsema}, R., {et~al.} 2018, \aap, 619, A9

\bibitem[{{Schmid} {et~al.}(2006){Schmid}, {Joos}, \& {Tschan}}]{Schmid2006}
{Schmid}, H.~M., {Joos}, F., \& {Tschan}, D. 2006, \aap, 452, 657

\bibitem[{{Schneider} {et~al.}(2018){Schneider}, {Debes}, {Grady},
  {G{\'a}sp{\'a}r}, {Henning}, {Hines}, {Kuchner}, {Perrin}, \&
  {Wisniewski}}]{Schneider2018}
{Schneider}, G., {Debes}, J.~H., {Grady}, C.~A., {et~al.} 2018, \aj, 155, 77

\bibitem[{{Schneider} {et~al.}(1999){Schneider}, {Smith}, {Becklin}, {Koerner},
  {Meier}, {Hines}, {Lowrance}, {Terrile}, {Thompson}, \&
  {Rieke}}]{Schneider1999}
{Schneider}, G., {Smith}, B.~A., {Becklin}, E.~E., {et~al.} 1999, \apjl, 513,
  L127

\bibitem[{{Schneider} {et~al.}(2009){Schneider}, {Weinberger}, {Becklin},
  {Debes}, \& {Smith}}]{Schneider2009}
{Schneider}, G., {Weinberger}, A.~J., {Becklin}, E.~E., {Debes}, J.~H., \&
  {Smith}, B.~A. 2009, \aj, 137, 53

\bibitem[{{Smart}(1930)}]{Smart1930}
{Smart}, W.~M. 1930, \mnras, 90, 534

\bibitem[{{Tazaki} {et~al.}(2016){Tazaki}, {Tanaka}, {Okuzumi}, {Kataoka}, \&
  {Nomura}}]{Tazaki2016}
{Tazaki}, R., {Tanaka}, H., {Okuzumi}, S., {Kataoka}, A., \& {Nomura}, H. 2016,
  \apj, 823, 70

\bibitem[{{Telesco} {et~al.}(2000){Telesco}, {Fisher}, {Pi{\~n}a}, {Knacke},
  {Dermott}, {Wyatt}, {Grogan}, {Holmes}, {Ghez}, {Prato}, {Hartmann}, \&
  {Jayawardhana}}]{Telesco2000}
{Telesco}, C.~M., {Fisher}, R.~S., {Pi{\~n}a}, R.~K., {et~al.} 2000, \apj, 530,
  329

\bibitem[{{Thalmann} {et~al.}(2011){Thalmann}, {Janson}, {Buenzli}, {Brandt},
  {Wisniewski}, {Moro-Mart{\'{\i}}n}, {Usuda}, {Schneider}, {Carson},
  {McElwain}, {Grady}, {Goto}, {Abe}, {Brandner}, {Dominik}, {Egner}, {Feldt},
  {Fukue}, {Golota}, {Guyon}, {Hashimoto}, {Hayano}, {Hayashi}, {Hayashi},
  {Henning}, {Hodapp}, {Ishii}, {Iye}, {Kandori}, {Knapp}, {Kudo}, {Kusakabe},
  {Kuzuhara}, {Matsuo}, {Miyama}, {Morino}, {Nishimura}, {Pyo}, {Serabyn},
  {Suto}, {Suzuki}, {Takahashi}, {Takami}, {Takato}, {Terada}, {Tomono},
  {Turner}, {Watanabe}, {Yamada}, {Takami}, \& {Tamura}}]{Thalmann2011}
{Thalmann}, C., {Janson}, M., {Buenzli}, E., {et~al.} 2011, \apjl, 743, L6

\bibitem[{{Thebault} \& {Kral}(2019)}]{Thebault2019}
{Thebault}, P. \& {Kral}, Q. 2019, arXiv e-prints, arXiv:1904.05395

\bibitem[{{Wahhaj} {et~al.}(2014){Wahhaj}, {Liu}, {Biller}, {Nielsen},
  {Hayward}, {Kuchner}, {Close}, {Chun}, {Ftaclas}, \& {Toomey}}]{Wahhaj2014}
{Wahhaj}, Z., {Liu}, M.~C., {Biller}, B.~A., {et~al.} 2014, \aap, 567, A34

\bibitem[{{Wyatt} {et~al.}(1999){Wyatt}, {Dermott}, {Telesco}, {Fisher},
  {Grogan}, {Holmes}, \& {Pi{\~n}a}}]{Wyatt1999}
{Wyatt}, M.~C., {Dermott}, S.~F., {Telesco}, C.~M., {et~al.} 1999, \apj, 527,
  918

\end{thebibliography}

\begin{appendix} 

\section{MCMC fit of the elllipse}
\label{App_MCMC}

\begin{figure}
 \centering
 \includegraphics[width=\hsize]{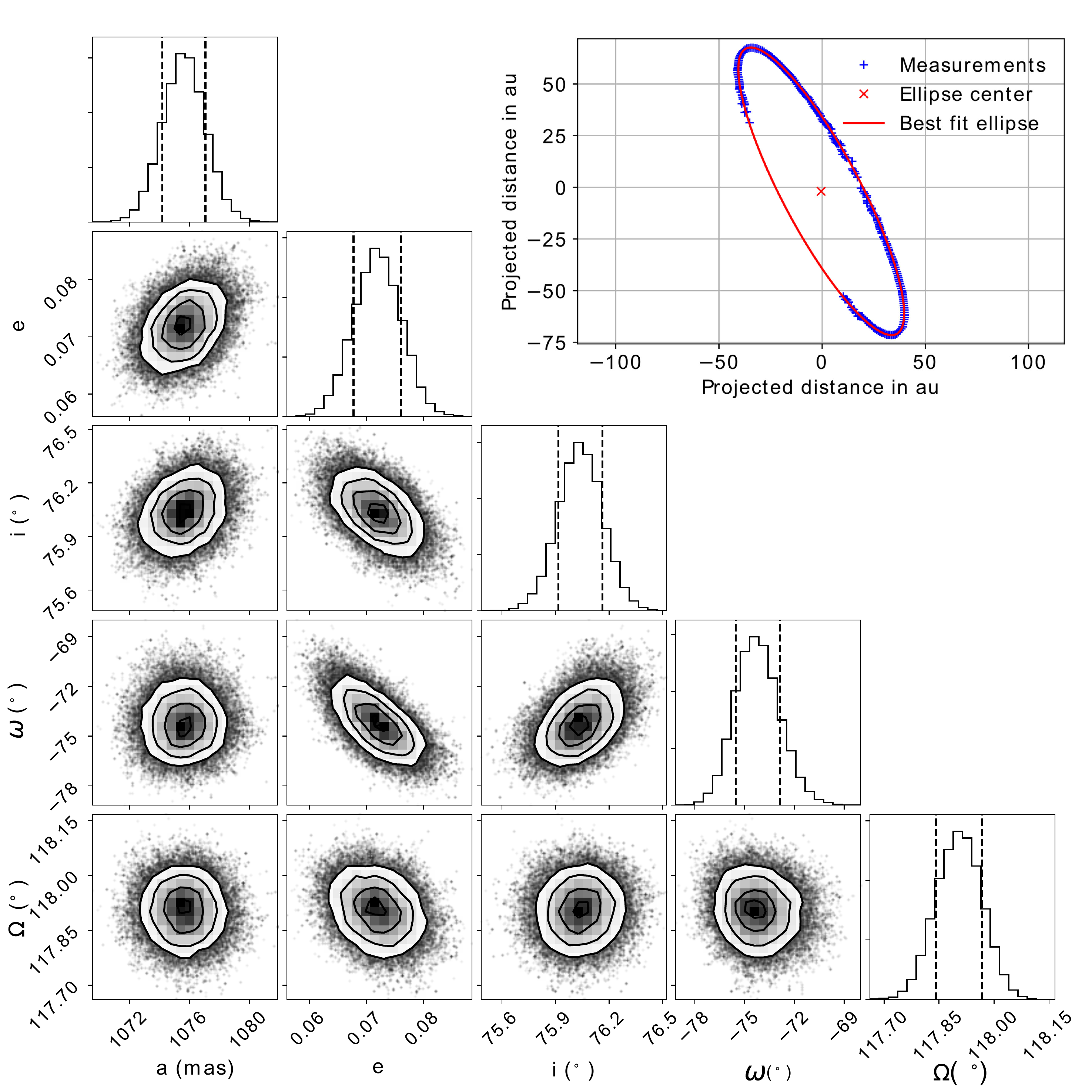}
 \caption{Marginal probability distribution of the deprojected elements of the ring. The top-right inset shows in blue the data points used as input for the fit, and in red the best ellipse.}
 \label{fig_corner_plot_MCMC}%
 \end{figure}

An ellipse is  characterised by the five following parameters: the centre coordinates ($x_0$,$y_0$), the semi-major and semi-minor axes $a$ and $b,$ and the position angle of the semi-major axis $PA$. \citet{Ray2008} proposed a geometric approach to characterise the goodness of fit between an ellipse parametrised by the model vector $\mathbf{u}=(x_0,y_0,a,b,PA)$ and a set of data points. We adopted their definition: if $F_i(\mathbf{u})$ is the distance between the $i^{th}$  data point and its projection on the ellipse as defined in their Fig. 3, the misfit is the sum of $F_i^2(\mathbf{u})$. Finding the minimum misfit is a non-linear least-square problem that we choose to solve with a Markov Chain Monte Carlo technique (MCMC). We implemented the affine-invariant ensemble sampler called emcee \citep{Foreman-Mackey2013}. We assumed uniform priors for each ellipse parameter. The posterior probability density function for the deprojected ellipse parameters is shown in Fig. \ref{fig_corner_plot_MCMC}, as well as the data points and best model ellipse (inset image).

 \section{Correcting the extracted phase function from the effect of convolution }
\label{App_convolution}

The size of the PSF and particularly the extension of the PSF wings have an impact because they dilute the flux of the ring. 
In absolute terms, it affects the polarised flux of the disc extracted from the image. In relative terms, it affects the pSPF extracted from the data because the impact of the convolution varies along the ring.
To correct it, we generated an unconvolved model of the HR\,4796A debris ring to reproduce the observations, then we convolved this model with the ZIMPOL PSF, a technique already applied in \citet{Rodigas2014} for this system. 

\begin{figure}
 \centering
 \includegraphics[width=\hsize]{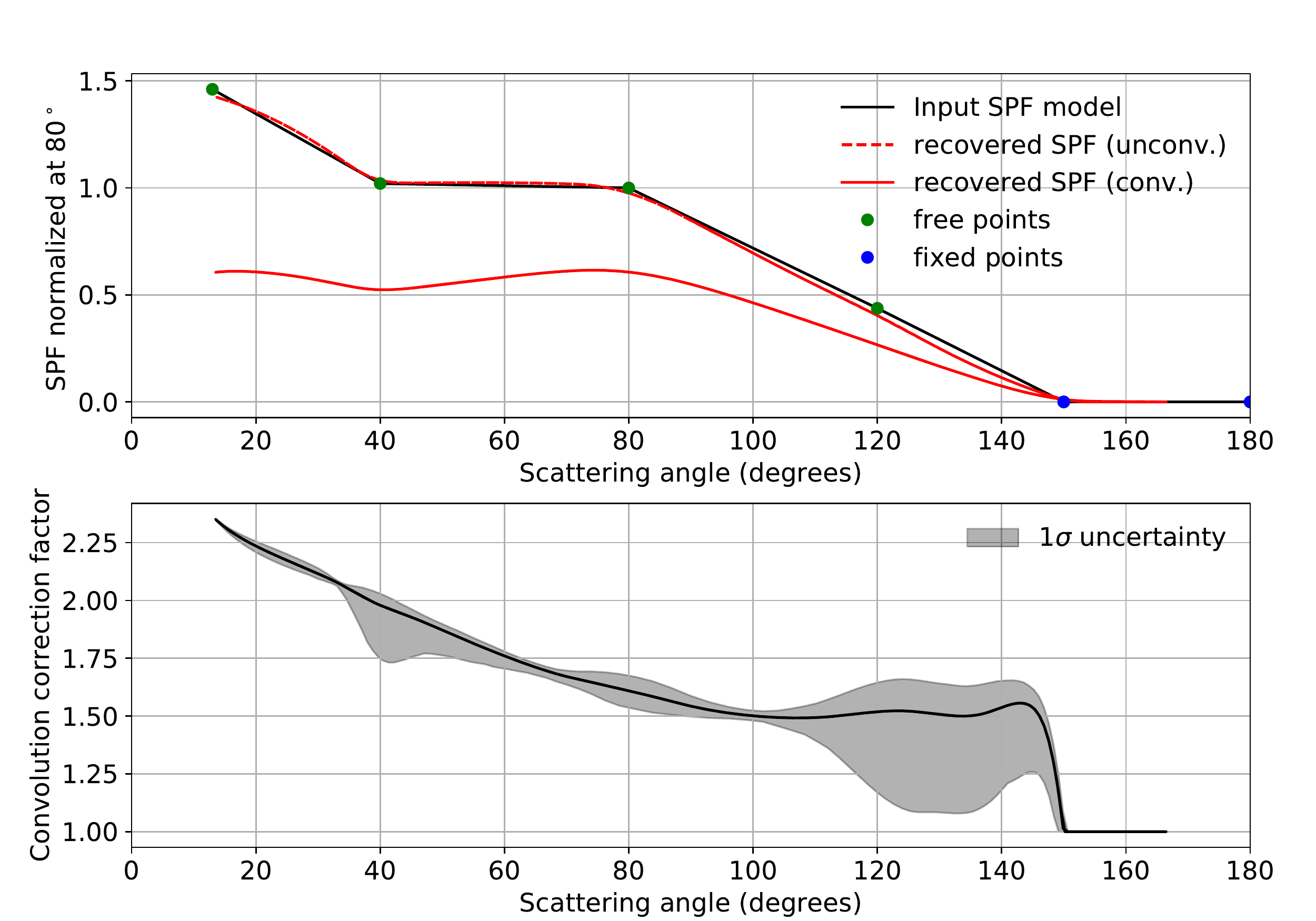}
 \caption{Top: Model of the best polarized SPF matching the data (black curve), described as a piecewise linear function  with 6 nodes (the 4 green dots are the degrees of freedom for the fit, the 2 blue dots are fixed parameters). This input pSPF is compared to the pSPF as extracted from the synthetic image using the elliptical aperture technique described in Sect. \ref{sec_SPF_extraction} before convolution (red dashed line) and after convolution (red plain line). Bottom: Correction factor account for the effect of convolution (corresponding to the ratio between the dashed and plain red line). The grey shaded area corresponds to the $1\sigma$ uncertainty propagated from the MCMC result.}
 \label{fig_convolution_model}
 \end{figure}

\begin{figure*}
 \centering
 \includegraphics[width=\hsize]{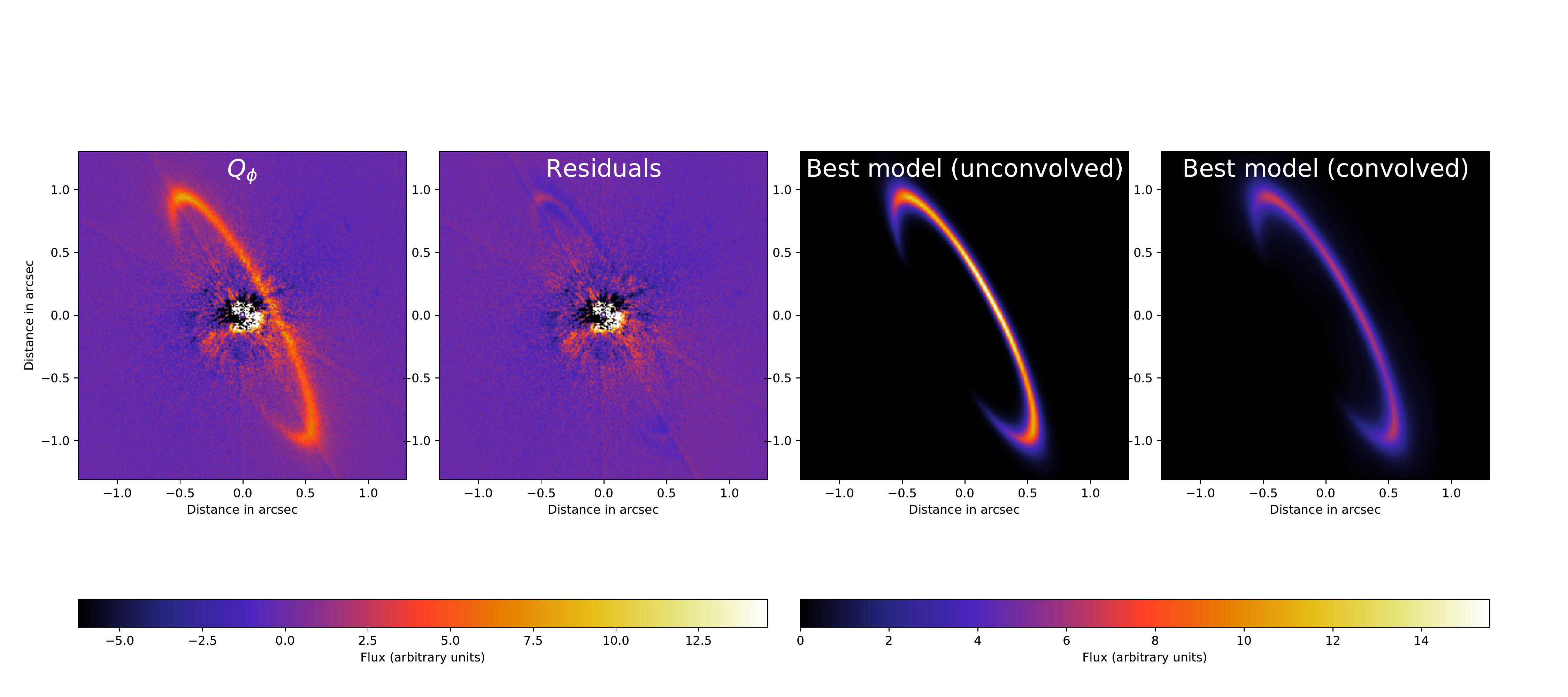}
 \caption{From left to right: Original $Q_\phi$ image, residuals after subtraction of the best model showing only some residual disc flux in the brighter north ansa, and the unconvolved model and the convolved model. The first two images and last two images have the same linear colour scale.}
 \label{fig_best_modell}%
 \end{figure*}

To find the best model reproducing the data, we used a python implementation\footnote{GraTeR was implemented in python as part of the high-contrast pipeline VIP \citep{Gomez2017}} of the GraTeR code \citep{Augereau1999}. The complete description of the input parameters is given in Appendix B of \citet{Milli2017}. We used here the geometrical parameters of the ring described in Table \ref{tab_ring_morphological}.  Regarding the vertical and radial dust density distribution, we used a reference scale height of $\xi_0$=1\,au at $a=77.4\,au$, a Gaussian vertical profile $\gamma=2$, a linear flaring $\beta=1$, and inner and outer radial density slopes $a_{in}=16.1$ and $a_{out}=-13.9$. They were optimised manually and fixed as we only want here a good model reproducing the data. We then used a custom phase function that we optimised to best reproduce the data. As we want to keep the number of free parameters as low as possible, we observed that the extracted phase function can be well approximated by a piecewise function as shown in Fig. \ref{fig_convolution_model} (top), containing five linear segments. Out of the six nodes defining those five linear segments, four are kept as free parameters, for scattering angles corresponding to $13^\circ$, $40^\circ$, $80^\circ$ and $120^\circ$. The last two nodes at $160^\circ$ and $180^\circ$ are set to zero as we have no constraints in this region.  
The free parameters of the fit are called $s_{13}$, $s_{40}$, $s_{80}$ and $s_{120}$ and correspond to the product of the pSPF function times the disc total scattering cross-section. The best model is shown in Fig. \ref{fig_best_modell}. It has a reduced $\chi^2$ of 1.18. We reproduced the best scattering phase function in Fig. \ref{fig_convolution_model} which shows both the input pSPF used in the model (black line) and the retrieved pSPF following our measurement procedure detailed in section \ref{sec_SPF_extraction} (red dashed line). This verification validates our pSPF extraction procedure as both curves agree very well. 

\begin{figure}
 \centering
 \includegraphics[width=\hsize]{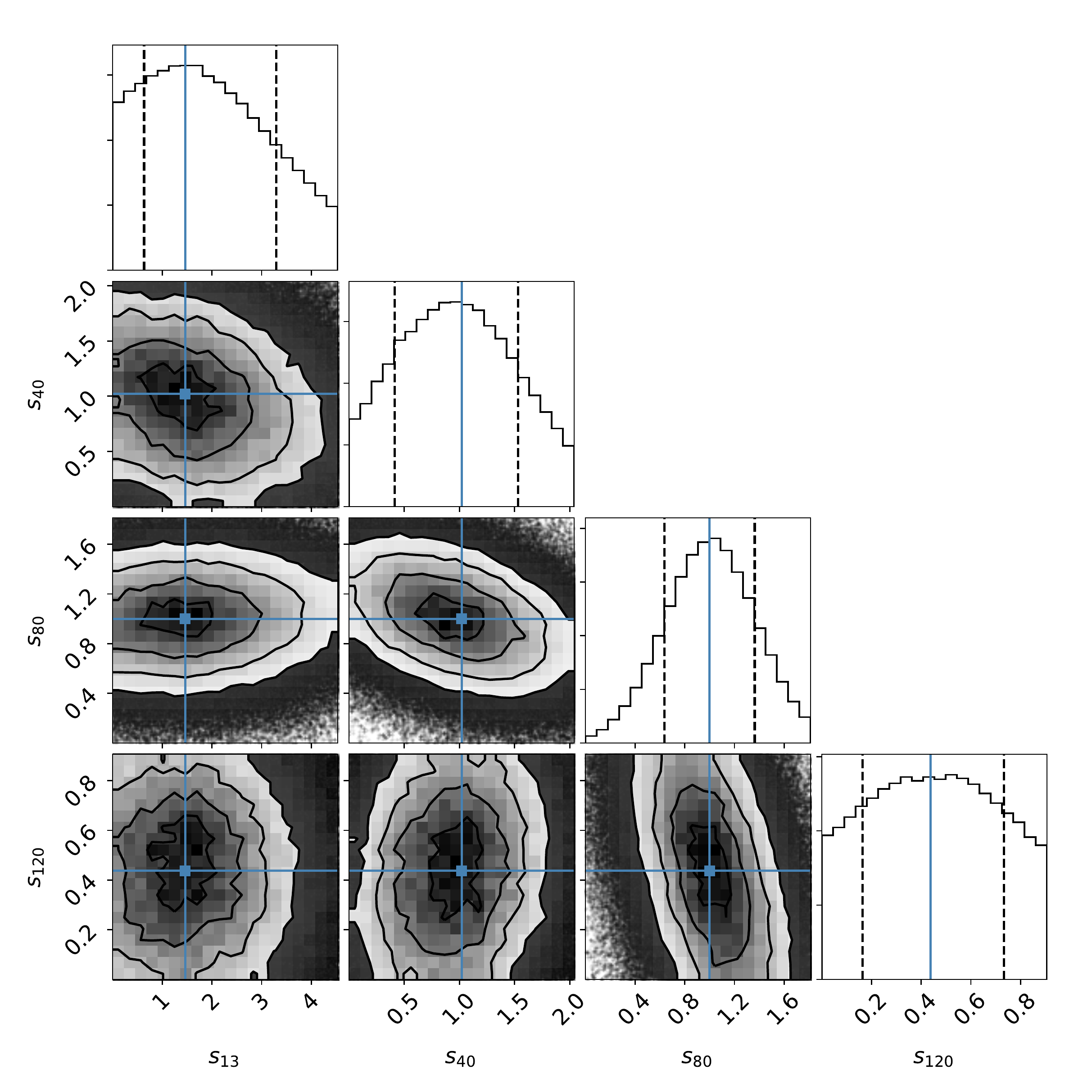}
 \caption{Marginal probability distribution of the 4 free parameters defining the pSPF.}
 \label{fig_corner_plot_model_scattered_light}
 \end{figure}

To derive the uncertainty on the convolution correction factor, we used once again the MCMC implementation of the affine-invariant ensemble sampler emcee. We used 120 walkers, a burning phase of 4,000 steps and then iterated over 16,000 steps for each of the walkers. The chain mean acceptance fraction was 0.5, and the maximum length for the auto-correlation time was 66. The posterior distributions of the four pSPF free parameters $s_{13}$, $s_{40}$, $s_{80}$ and $s_{120}$ are shown in in Fig. \ref{fig_corner_plot_model_scattered_light}. We then propagated the uncertainty in those parameters in the uncertainty on the correction factor from the convolution. This is shown in the grey shaded area in Fig. \ref{fig_best_modell} (bottom). This uncertainty was combined together with the other sources of uncertainty while extracting the pSPF.


 \section{Mie and DHS models}
\label{App_best_model}

We summarise in Table \ref{tab_chi2_Mie_DHS} the properties of the Mie or DHS models found to best describe the SED  \cite[][]{Milli2015}, the spectral reflectance, the unpolarised SPF below $45^\circ$ scattering angle \citep{Milli2017}, and the pSPF described in this work. As explained in Sect. \ref{sec_Mie}, the models best matching the pSPF have a minimum particle size of 0.1\micron, which is neither compatible with the SED nor with the spectral reflectance. 

\begin{table}
\caption{Goodness-of-fit estimates and corresponding parameters for the best models with respect to the SED or the scattered light observables.}            
\centering          
\label{tab_chi2_Mie_DHS}      
\begin{tabular}{ c | c c c c }     
\hline\hline       
  & best & best    & best SPF & best \\
  & SED\tablefootmark{a} & reflectance\tablefootmark{b} & $\varphi \leq 45^\circ$\tablefootmark{b} & polarised SPF\\
  \hline                  
Theory               & DHS    & Mie & Mie & Mie\\
$\nu$                 & -3.5     & -3.5 & -5.5  & -3.5 \\
$q_\text{Sior}$  & 0.2       & 0.2 & 0.0 & 1.\\
$p_{H_2O}$      & 3.1\%   & 1.0\% & 90\% & 90\% \\
$s_\text{min}$  & 1.78      & 0.3 & 17.8 & 0.1 \\
P                       & 20.0\%      & 40.0\% & 0.10\% & 80\%\\
\hline                  
$\chi^2_\text{SED}$                     & \textbf{1.7} & 129                & 48       & 84.6  \\
$\chi^2_\text{refl.}$                      & 352             & \textbf{1.4}  & 6.2          & 35.2  \\
$\chi^2_\text{SPF}$                     & 394             & 442               & \textbf{3.8} &  19.3 \\
$\chi^2_\text{pSPF}$                     &       14.1              &        16.9            &      12.2                 &  \textbf{2.4}  \\
\hline
\end{tabular}
\tablefoot{
\tablefoottext{a}{Best model explaining the SED already presented in \citet{Milli2015} and displayed here as a reference.}
\tablefoottext{b}{Best model explaining the spectral reflectance and SPF already presented in \citet{Milli2017} and displayed here as a reference.}
}
\end{table}

\end{appendix}
 
\end{document}